\documentclass[12pt,a4paper]{article}
\usepackage{amssymb}
\usepackage{amsmath}
\usepackage{amsfonts}
\usepackage{epsfig}
\usepackage{feynmf}
\usepackage{color}
\usepackage{hyperref}

\usepackage{graphicx}
	\graphicspath{{./Figs/}}

\usepackage[margin=0pt, labelfont={bf,footnotesize}, font=small, justification=raggedright]{caption}

\usepackage{bm}
\outer\def\beginsection#1\par{\medbreak\bigskip
      \message{#1}\leftline{\bf#1}\nobreak\medskip
\vskip-\parskip
      \noindent}

\newcommand{\eq}{\begin{equation}}
\newcommand{\eqx}{\end{equation}}
\newcommand{\eqn}{\begin{eqnarray}}
\newcommand{\eqnx}{\end{eqnarray}}
\newcommand{\bi}{\begin{itemize}}
\newcommand{\ei}{\end{itemize}}

\def\a{\alpha}

\def\Ord{{\cal O}}

\def\be{\begin{equation}}
\def\ee{\end{equation}}
\def\ba{\begin{eqnarray}}

\newcommand{\ea}[1]{\begin{align} #1 \end{align}}
\newcommand{\sea}[1]{\begin{subequations}\begin{align} #1 \end{align}\end{subequations}}
\newcommand{\seal}[2]{\begin{subequations}\label{#1} \begin{align} #2 \end{align}\end{subequations}}
\newcommand{\seq}[1]{\begin{equation} \begin{split} #1 \end{split} \end{equation}}

\newcommand{\LN}[2]{\log | z_{#1}-z_{#2} |}
\newcommand{\zz}[2]{z_{#1}-z_{#2}}
\newcommand{\zbzb}[2]{\bar{z}_{#1}-\bar{z}_{#2}}
\newcommand{\aqk}[1]{ \alpha' q k_{#1}}
\newcommand{\qk}[1]{ q k_{#1}}
\newcommand{\teq}[1]{\theta_{#1}\epsilon_{#1} q}
\newcommand{\temu}[1]{(\theta_{#1}\epsilon_{#1}^\mu)}

\newcommand{\tbeq}[1]{\bar{\theta_{#1}}\bar{\epsilon}_{#1} q}

\newcommand{\tbenu}[1]{\bar{\theta_{#1}}\bar{\epsilon}_{#1}^\nu}

\usepackage{empheq}

\usepackage[titles]{tocloft}

\usepackage{titlesec}
\titleformat*{\section}{\large  \bfseries }
\titleformat*{\subsection}{\normalsize  \bfseries }

\begin{document}
\begin{titlepage}
\hfill \hbox{NORDITA-2016-025}
\vskip 1.5cm
\vskip 1.0cm
\begin{center}
{\Large \bf  Subsubleading soft theorems of gravitons and dilatons in the bosonic string}
 
\vskip 1.0cm {\large Paolo
Di Vecchia$^{a,b}$,
Raffaele Marotta$^{c}$, Matin Mojaza$^{b}$,} \\[0.7cm] 
{\it $^a$ The Niels Bohr Institute, University of Copenhagen,\\
Blegdamsvej 17, DK-2100 Copenhagen \O , Denmark}\\[.2cm] 
{\it $^b$ NORDITA, KTH Royal Institute of Technology and Stockholm 
University, \\
Roslagstullsbacken 23, SE-10691 Stockholm, Sweden}\\[.2cm]
{\it $^c$  Istituto Nazionale di Fisica Nucleare, Sezione di Napoli. \\
Complesso Universitario di Monte S. Angelo ed. 6, via Cintia, 80126, Napoli, Italy
}
\end{center}
\begin{abstract}

Starting from the amplitude with an arbitrary number of massless closed states of the bosonic
string, we compute the soft limit when one of the states becomes soft to
subsubleading order in the soft momentum expansion, and we show that when
the soft state is a graviton or a dilaton, the full string amplitude can be
expressed as a soft theorem through subsubleading order. It turns out that
there are string corrections to the field theoretical limit in the case of a
soft graviton, while for a soft dilaton the string corrections vanish.
We then show that the new soft theorems, including the string corrections, can be 
simply obtained 
from the exchange diagrams where the soft state is attached to the other external states through the three-point string vertex of three massless states. In the soft-limit, the propagator of the exchanged state is divergent, and at tree-level these are the only divergent contributions to the full amplitude. However, they do not form a gauge invariant subset and must be supplemented with extra non-singular terms.
The requirement of gauge invariance then fixes the complete amplitude
through subsubleading order in the soft expansion, reproducing exactly what
one gets from the explicit calculation in string theory. From this it is seen that the string corrections at subsubleading order arise as a consequence of the three-point amplitude having string corrections in the bosonic string.
When specialized to a soft dilaton, it remarkably turns out that the string corrections vanish and that the non-singular piece of the subsubleading term of the dilaton soft theorem is the generator of space-time special conformal transformation.

\end{abstract}
\end{titlepage}

\tableofcontents

\section{Introduction and summary of results}
\label{intro}

Tremendous progress is happening in understanding the soft factorizing 
behavior of scattering amplitudes and their relation to underlying, sometimes 
hidden, symmetries.
Most remarkable, perhaps, are the suggestions that the soft behavior of 
particularly the gravity and Yang-Mills S-matrices are related to asymptotic 
symmetries in general relativity and in gauge theories~\cite{asymp}.
Also remarkable are the similarities pointed out very recently between the soft 
behavior of the gravity/string dilaton and the Ward identities of scale and 
special conformal transformations~\cite{Boels:2015pta,DiVecchia:2015jaq}.
New uses of soft theorems are also being discovered in the more modern field of amplitudes~\cite{Cheung:2014dqa,Luo:2015tat}.

Soft theorems, however, have a long history, and go back to the seminal works 
in the 1950s on low-energy photon scattering~\cite{Low} and in the 1960s on 
soft-graviton scattering~\cite{Weinberg}, when they were realized to be 
important consequences of gauge invariance. Discussions on the generic 
subleading behavior of soft gluon and graviton scattering were recently 
taken up in~\cite{GenericSubStart}, and has, since the suggested relations 
to asymptotic symmetries~\cite{asymp}, received enormous attention, not 
only in gravity and Yang-Mills 
theory~\cite{SoftGravityYangMills,BDDN,BianchiR2phi}, but also 
their extensions in supersymmetric theories~\cite{softsusy}, and in 
string theory~\cite{softstring,DiVecchia:2015oba}.
Double-soft theorems are also receiving increasing 
interest~\cite{DoubleSoft}, due to their potential to uncover hidden 
symmetries of the S-matrix (see e.g. Ref.~\cite{ArkaniHamed:2008gz} 
for a discussion on Adler's zeroes and the pion double-soft theorem).

Soft theorems in string theory 
were first discussed in the 1970s by
Ademollo et al.~\cite{Ademollo:1975pf} and by Shapiro~\cite{Shapiro:1975cz}
for tree diagram scattering amplitudes involving massless particles only, and 
with  particular emphasis on the string-dilaton as the soft-state (see also 
Refs.~\cite{YoneyaHata} for this study in string field theory).
In a recent work~\cite{DiVecchia:2015oba}, we have revived this line of 
studies by computing the soft behavior up to
subsubleading order, 
when a soft massless closed string-state is scattering 
on external tachyons. It turns out that this amplitude is determined by the 
same gauge invariance that also determines the soft-graviton behavior up 
to subsubleading order in field theory, derived in Ref.~\cite{BDDN}. 
Furthermore, we computed the leading soft-behavior 
of the antisymmetric Kalb-Ramond tensor in the scattering on other 
massless closed string states. At the same time we rederived the known 
results involving instead a soft graviton or dilaton, and showed by invoking a slight generalization of the analysis done in Ref.~\cite{BDDN} that the
leading soft behavior for both of them is again determined by field theory 
gauge invariance.
The aim of this work is to extend our previous analysis in the bosonic 
string to the subsubleading order for the case of a soft graviton or 
dilaton scattering on other closed massless states.
At this order, string corrections to the corresponding field theory soft 
theorems are expected to appear for the first time~\cite{BianchiR2phi}, and 
indeed this is what we find.  Their presence is  also expected   in heterotic   
string and it  is  due to  interaction terms of the type $\phi R^2$ which 
appear, to  order $\alpha'$, in the  effective actions of such string  
theories~\cite{Metsaev:1987zx}. String corrections to the  graviton 
soft theorem have  also been  computed in \cite{1512.00803}  in the 
case of  four point bosonic string  amplitudes.    We have extended  
this  analysis to $n+1$ point amplitudes   with a soft  dilaton or  
graviton and $n$ massless hard particles finding that  only the  
graviton soft operator is  modified by $\alpha'$-corrections.  
The lack of string corrections
in the soft behavior of the dilaton   
could be a signal  that the dilaton soft theorem  is  a consequence of   
some Ward identity as it  occurs   for the   Nambu-Goldstone boson of 
the spontaneously broken conformal invariance~\cite{DiVecchia:2015jaq}.  
 The similarities between these two  particles, both called dilaton, indeed 
deserve  a further study.   

Let us summarize our primary results before going through the calculational details.

In Ref.~\cite{DiVecchia:2015jaq} it has been  
shown that the field theory  amplitude 
for a soft graviton 
or a soft dilaton of soft momentum $q$, with $n$ other hard gravitons and/or dilatons can be written in the following factorized form:
 \ea{
M_{n+1}(q; k_i) \equiv \epsilon_{\mu \nu}^S M^{\mu \nu} (k_i; q)  
&= \kappa_D \left ( \hat{S}_q^{(-1)}+\hat{S}_q^{(0)}+\hat{S}_q^{(1)} \right ) 
M_n(k_i) + \Ord(q^2) \, , 
\label{softbe}
}
where $\kappa_D$ is related to the gravitational constant in $D$ space-time dimension, the superscript of each $\hat{S}_q^{(m)}$ indicates the order $m$ in 
$q$ of each term, and $M_n$ is the amplitude without the soft particle. 
$\epsilon_{\mu \nu}^S$ is the polarization of either the graviton or the
dilaton, which is symmetric under the exchange of $\mu$ and $\nu$.
In Ref.~\cite{DiVecchia:2015oba}, the above soft theorem to subleading order was shown to hold also in the framework of the bosonic string including also the Kalb-Ramond antisymmetric field both in the role of the soft state and as hard states.

In the cases of a soft graviton or dilaton, the first two terms are given 
by~\cite{DiVecchia:2015oba}:
\ea{
\hat{S}_q^{(-1)} &= \epsilon_{\mu \nu}^S \sum_{i=1}^n \frac{k_i^\mu k_i^\nu}{k_i \cdot q}
\ , \quad
\hat{S}_q^{(0)} =  \epsilon_{\mu \nu}^S\left ( - \frac{i q_\rho}{2} \right )
\sum_{i=1}^n \frac{k_i^\mu J_i^{\nu \rho} + k_i^\nu J_i^{\mu \rho}}{k_i 
\cdot q} \, ,
\label{leasublea}
}
where
\begin{eqnarray}
J_i^{\mu \nu} = L_i^{\mu \nu} + 
\mathcal{S}_i^{\mu \nu} \, , \quad
\mathcal{S}_i^{\mu \nu} = S_i^{\mu \nu} 
+ {\bar{S}}^{\mu \nu}_i 
\ ,
\label{JLS1}
\end{eqnarray}
\ea{
L_i^{\mu\nu} =i\left( k_i^\mu\frac{\partial }{\partial k_{i\nu}} -k_i^\nu
\frac{\partial }{\partial k_{i\mu}}\right) ,  \ 
S_i^{\mu\nu}=i\left( \epsilon_i^\mu\frac{\partial }{\partial \epsilon_{i\nu}} -
\epsilon_i^\nu\frac{\partial }{\partial \epsilon_{i\mu}}\right)  , \  
{\bar{S}}^{\mu\nu}_i=i\left( {\bar{\epsilon}}_i^\mu\frac{\partial }{\partial 
{\bar{\epsilon}}_{i\nu}} -{\bar{\epsilon}}_i^\nu\frac{\partial }{\partial 
{\bar{\epsilon}}_{i\mu}}\right)   .
\label{LandS}
}
while the third term was computed in the field theory limit in 
Ref.~\cite{DiVecchia:2015jaq}.  The method used is an extension of the one
of  Ref.~\cite{BDDN} and  the soft behavior in Eq.~(\ref{softbe}) is shown to 
be a direct consequence of the gauge invariance conditions 
\begin{eqnarray}
q_\mu M^{\mu \nu} (k_i; q) =q_\nu M^{\mu \nu} (k_i; q) =0 \, .
\label{gaugeinvcond}
\end{eqnarray}
In this paper we extend the previous method  to include string corrections 
and we check the final result by performing a direct calculation of 
the subsubleading term in the soft limit in the amplitude of the bosonic string
involving an arbitrary number of massless closed strings. This calculation
is performed by extending the technique
developed in Ref.~\cite{DiVecchia:2015oba} for the computation of the 
subleading term.  
As a result we obtain the following subsubleading term:
\seq{
S_q^{(1)} = &- \frac{\epsilon_{\mu \nu}^S}{2} \sum_{i=1}^n
\left [
\frac{q_\rho J_i^{\mu \rho} q_\sigma J_i^{\nu \sigma}}{k_i \cdot q}
+ \left (
\frac{k_i^\mu q^\nu}{k_i \cdot q} q^\sigma + q^\mu \eta^{\nu \sigma}- 
\eta^{\mu \nu} q^\sigma \right )
\frac{\partial}{\partial k_{i}^\sigma} \right .
 \\
&-\left (
\frac{q_\rho q_\sigma \eta_{\mu \nu} - q_\sigma q_\nu \eta_{\rho \mu} - 
q_\rho q_\mu \eta_{\sigma \nu}}{ k_i\cdot q}
\right )  \Pi_i^{\rho \sigma} 
 \\
&\left .  - \alpha' \left (q_\sigma k_{i\nu} \eta_{\rho \mu}+q_\rho k_{i\mu }
\eta_{\sigma \nu} - \eta_{\rho\mu}\eta_{\sigma \nu} (k_i \cdot q) - q_\rho 
q_\sigma \frac{k_{i\mu}k_{i\nu}}{k_i \cdot q} \right )
\Pi_i^{\rho \sigma}   \right ]  ,
\label{generalsubsub1}
}
where
\ea{
\Pi_i^{\rho \sigma} = 
\epsilon_i^\rho\frac{\partial }{\partial \epsilon_{i\sigma}} +
{\bar{\epsilon}}_i^\rho\frac{\partial }{\partial {\bar{\epsilon}}_{i\sigma}} \, .
\label{sumpoloperator1}
}
Only the symmetric part $\Pi_i^{\{\rho, \sigma\}} = 
\frac{\Pi_i^{\rho \sigma} + \Pi_i^{\sigma \rho}}{2}$ contributes in the previous 
expression because the polarization tensor 
$\epsilon^S_{\mu \nu}$ is symmetric in the indices $\mu$ and $\nu$.
The first two lines of Eq.~(\ref{generalsubsub1}) agree with the expression 
already presented in 
Ref.~\cite{DiVecchia:2015jaq},  while  the third line gives the string corrections. 

By choosing the polarization of the graviton, from
  Eqs.~(\ref{leasublea}) and (\ref{generalsubsub1}) we get the soft theorem 
of a graviton:
\seq{
M_{n+1}^{\rm graviton} = &\, \kappa_D \, \epsilon_{\mu \nu}^{g} \sum_{i=1}^n  \left[ 
\frac{k_i^\mu k_i^\nu - i q_\rho k_i^\mu J_{i}^{\nu \rho} - \frac{1}{2}
q_\rho J_i^{\mu \rho} q_\sigma J_i^{\nu \sigma}}{k_i q}  \right.   \\
& \left .  - \frac{\alpha'}{2} \left (q_\sigma k_{i\nu} 
\eta_{\rho \mu}+q_\rho k_{i\mu }
\eta_{\sigma \nu} - \eta_{\rho\mu}\eta_{\sigma \nu} (k_i \cdot q) - 
q_\rho q_\sigma \frac{k_{i\mu}k_{i\nu}}{k_i \cdot q} \right )
\Pi_i^{\{\rho, \sigma\}}  
\right] M_n \, , 
\label{gravisoftfina}
}
while,
by choosing the polarization tensor of the dilaton $\epsilon^{S}_{\mu\nu} \to
\frac{1}{\sqrt{D-2}}\left(
\eta_{\mu \nu} - q_\mu {\bar{q}}_\nu - q_\nu {\bar{q}}_\mu\right)$, 
we  get the soft theorem  for a dilaton:
\seq{
M_{n+1}^{\rm dilaton} = &\,
\frac{\kappa_D}{\sqrt{D-2}} \left[ 2 - \sum_{i=1}^n k_{i\mu} 
\frac{\partial}{\partial k_{i\mu}} 
\right .  \\
& \left. + \frac{1}{2}  \sum_{i=1}^n
\left( q^\rho {\hat{K}}_{i \rho} 
 + \frac{q^\rho q^\sigma}{k_i q}
\left(
\mathcal{S}_{i, \rho \mu}\eta^{\mu \nu} \mathcal{S}_{i \nu \sigma} + D
\Pi_{i,\{\rho, \sigma\}} \right) 
\right)  \right] M_n \, , 
\label{dilasoftfin}
}
where
\begin{eqnarray}
&&  {\hat{K}}_{i\mu} = 2 \left[ \frac{1}{2} k_{i \mu} \frac{\partial^2}{\partial
k_{i\nu} \partial k_i^\nu} 
-k_{i}^{\rho} \frac{\partial^2}{\partial
k_i^\mu \partial k_{i}^{\rho}} 
+ i \mathcal{S}_{i,\rho \mu} \frac{\partial}{\partial k_i^\rho}  \right] \, .
\label{hatDhatKmu1}
\end{eqnarray}
Remarkably, these operators are nothing but the generators of space-time special conformal transformations acting in momentum space.
As recently shown in Ref.~\cite{DiVecchia:2015jaq}, these operators also control the soft behavior of the Nambu-Goldstone bosons of spontaneously broken conformal invariance and an interesting application of this recently appeared in Ref.~\cite{Luo:2015tat}.
It would be interesting to understand the physical reason for why these 
generators appear in the soft limit of the string dilaton.
Notice furthermore that the string corrections vanish completely for a soft dilaton.

The paper is organized as follows. In Sect.~\ref{stringdilaton} we write the
amplitude with an arbitrary number of massless states of the closed bosonic string and we
perform the limit in which one of them (a dilaton or a graviton) becomes soft.
In Sect.~\ref{softdilasoftgravi} we derive the explicit form of the soft behavior
for a dilaton and for a graviton and we give a physical interpretation of the
various terms that appear. In Sect.~\ref{gaugeinvariance1} we derive the string 
corrections to the soft theorem through gauge invariance from the string corrections of the 
three-point amplitude for massless closed string states. Finally, details 
of the calculations presented in Sect.~\ref{stringdilaton} are given in two 
Appendices.

\section{Amplitude of one soft and $n$ massless closed strings}
\label{stringdilaton}

In this section we consider the amplitude with $n+1$ massless closed string 
states and we study its behavior in the limit in which one of the massless states 
is soft. We start by summarizing the results presented in 
Ref.~\cite{DiVecchia:2015oba}, where more details may be found.

The amplitude involving $n+1$ massless closed string states can be written as
\ea{
 M_{n+1} \sim
 &
 \int \frac{\prod_{i=1}^n d^2z_i\,d^2 z}{dV_{abc}}  \int d \theta  
 \prod_{i=1}^n d\theta_i ~ \langle 0| e^{i( \theta \epsilon^\mu_q \partial_{z}+ \sqrt{\frac{\alpha'}{2}} q^\mu)X_\mu(z)}~\prod_{i=1}^ne^{i( \theta_i \epsilon_i^{\mu_i} \partial_{z_i}+ \sqrt{\frac{\alpha'}{2}} k^{\mu_i}_i)X_{\mu_i}(z_i)}|0 \rangle \nonumber\\
&\times \int d {\bar{\theta}}  
\prod_{i=1}^{n} d {\bar{\theta}}_i \langle 0|  e^{i( \bar{\theta} \bar{\epsilon}^\mu_q \partial_{\bar{z}}+ \sqrt{\frac{\alpha'}{2}} q^\mu)X_\mu(\bar{z})}~\prod_{i=1}^ne^{i( \bar{\theta}_i \bar{\epsilon}_i^{\nu_i} \partial_{\bar{z}_i}+ \sqrt{\frac{\alpha'}{2}} k^{\nu_i}_i)X_{\nu_i}(\bar{z}_i)} |0 \rangle \ .
\label{amplitheta}
}
where we assume that {$\theta, \bar{\theta}, \epsilon, \bar{\epsilon}$} are Grassmann variables, 
and we use the definition 
\mbox{$\epsilon_{i \,  \mu \nu} \equiv \epsilon_{i \, \mu} 
{\bar{\epsilon}}_{i \, \nu}$} for the polarization tensor. We consider 
the soft string to be the one with momentum $q$ and polarization 
$\epsilon_{q, \mu \nu}$.
After using the contraction $\langle X^\mu (z) X^{\nu} (w) \rangle = - 
\eta^{\mu \nu} \log (z-w)$ and performing the integration over the 
Grassmann variables $\theta$ and ${\bar{\theta}}$, the expression 
reduces to a form which can formally be written in two parts:
\begin{eqnarray}
M_{n+1} =   M_n * S \ ,
\label{MMS}
\end{eqnarray}
where by $*$ a convolution of integrals is understood, and where
\seq{
S \equiv  
&\, 
\kappa_D
\int \frac{d^2 z}{2 \pi} \,\, \sum_{i=1}^{n} \left(\theta_i \frac{ (\epsilon_q \epsilon_i)}{(z-z_i)^2} +    \sqrt{\frac{\alpha'}{2}} \frac{(\epsilon_q k_i)}{z-z_i} \right) \sum_{j=1}^{n} \left({\bar{\theta}}_j \frac{ ({\bar{\epsilon}}_q {\bar{\epsilon}}_j)}{({\bar{z}}- {\bar{z}}_j)^2} +    \sqrt{\frac{\alpha'}{2}} \frac{({\bar{\epsilon}}_q k_i)}{{\bar{z}}-{\bar{z}}_i} \right)  \\
& \times \exp \left[ - \sqrt{\frac{\alpha'}{2}}  \sum_{i=1}^{n} \theta_i \frac{(\epsilon_i q)  }{z-z_i} \right] \exp \left[ - \sqrt{\frac{\alpha'}{2}}  \sum_{i=1}^{n} {\bar{\theta}}_i \frac{({\bar{\epsilon}}_i q)  }{{\bar{z}}-{\bar{z}}_i} \right]\prod_{i=1}^{n} |z- z_i|^{\alpha' q k_i} \, ,
\label{last3lines}
}
is the part describing the soft particle, and
\seq{
M_n = &\, \frac{8\pi}{\alpha'}\left (\frac{\kappa_D}{2\pi}\right )^{n-2} 
 \int \frac{\prod_{i=1}^n d^2z_i }{dV_{abc}} \int \left[\prod_{i=1}^n d\theta_i \prod_{i=1}^{n} d {\bar{\theta}}_i \right]   \prod_{i<j} |z_i - z_j |^{\alpha' k_i k_j}   \\
& \times \exp \left[ -\sum_{i<j}  \frac{\theta_i \theta_j}{(z_i - z_j)^2}  (\epsilon_i \epsilon_j) + \sqrt{\frac{\alpha'}{2}} \sum_{i \neq j} \frac{ \theta_i (\epsilon_i k_j) }{z_i - z_j}  \right]  \\
& \times \exp \left[- \sum_{i<j}  \frac{{\bar{\theta}}_i {\bar{\theta}}_j}{({\bar{z}}_i - {\bar{z}}_j)^2}  ({\bar{\epsilon}}_i {\bar{\epsilon}}_j) + \sqrt{\frac{\alpha'}{2}} \sum_{i \neq j} \frac{ {\bar{\theta}}_i ({\bar{\epsilon}}_i k_j) }{{\bar{z}}_i - {\bar{z}}_j}  \right] ,
\label{nonsoftonly}
}
is the amplitude of $n$ massless states without the soft particle.

We eventually want to find a soft operator $\hat{S}$ such that \mbox{$\hat{S}M_n = M_n \ast S$} through order $q^1$.
This can be done by expanding $S$ for small $q$ and keep terms in the integrand up to the order $q^2$, since higher orders of the integrand cannot yield terms of order $q^1$ after integration. 
It is useful then to divide $S$ in three parts:
\begin{eqnarray}
S = \kappa_D \left ( S_1 + S_2 + S_3 \right ) + \Ord(q^2) \ , 
\label{SSi}
\end{eqnarray}
defined by:
\seq{
S_{1} = &\,
\frac{\alpha'}{2} \int \frac{ d^2 z}{2\pi}\, \sum_{i=1}^{n}\frac{(\epsilon_q k_i)}{z-z_i}\sum_{j=1}^{n}  \frac{({\bar{\epsilon}}_q k_j)}{{\bar{z}}-{\bar{z}}_j} \prod_{i=1}^{n} |z- z_i|^{\alpha' q k_i}
\\
& \times 
\Bigg\{ 
1
 - \sqrt{\frac{\alpha'}{2}}
\sum_{k=1}^{n}  \Bigg(
\theta_k  \frac{(\epsilon_k q)  }{z-z_k} + {\bar{\theta}}_k \frac{({\bar{\epsilon}}_k q)  }{{\bar{z}}-{\bar{z}}_k}
\Bigg) 
+ \frac{1}{2}\left( \frac{\alpha'}{2}\right )
\\
&
\times\Bigg[
\left( \sum_{h=1}^{n}   \theta_h  \frac{(\epsilon_h q)  }{z-z_h}  \right)^2
+
\left( \sum_{h=1}^{n}   {\bar{\theta}}_h \frac{({\bar{\epsilon}}_h q)  }{{\bar{z}}-{\bar{z}}_h}   \right)^2
+  
2\left( \sum_{h=1}^{n}   \theta_h  \frac{(\epsilon_h q)  }{z-z_h}  \right) \left( \sum_{h=1}^{n}   {\bar{\theta}}_h \frac{({\bar{\epsilon}}_h q)  }{{\bar{z}}-{\bar{z}}_h}   \right)
\Bigg]
\Bigg\} \, ,
\label{S1}
}
\seq{
S_2 = &\,
\int \frac{ d^2 z}{2\pi} \sum_{i=1}^{n} \left(\theta_i \frac{ (\epsilon_q 
\epsilon_i)}{(z-z_i)^2}\right) 
\sum_{j=1}^{n} \left({\bar{\theta}}_j \frac{ ({\bar{\epsilon}}_q 
{\bar{\epsilon}}_j)}{({\bar{z}}- {\bar{z}}_j)^2}  \right)\prod_{\ell=1}^{n}
 |z- z_{\ell}|^{\alpha' q k_{\ell}}
 \\
& \times \Bigg\{
1
-  
\sqrt{\frac{\alpha'}{2}}  \sum_{k=1}^{n} \Bigg( 
\theta_k \frac{\epsilon_k  q}{z-z_k}
+
{\bar{\theta}}_k \frac{({\bar{\epsilon}}_k q)  }{{\bar{z}}-{\bar{z}}_k}
 \Bigg)
 + \frac{1}{2}\left( \frac{\alpha'}{2}\right )
\\
& 
\times \Bigg[
\left( \sum_{h=1}^{n}   \theta_h  \frac{(\epsilon_h q)  }{z-z_h}  \right)^2
+
\left( \sum_{h=1}^{n}   {\bar{\theta}}_h \frac{({\bar{\epsilon}}_h q)  }{{\bar{z}}-{\bar{z}}_h}   \right)^2
+  
2\left( \sum_{h=1}^{n}   \theta_h  \frac{(\epsilon_h q)  }{z-z_h}  \right) \left( \sum_{h=1}^{n}   {\bar{\theta}}_h \frac{({\bar{\epsilon}}_h q)  }{{\bar{z}}-{\bar{z}}_h}   \right)
\Bigg]  \Bigg\}\, ,
\label{S2}
}
\seq{
S_3 = &\,
\sqrt{\frac{\alpha'}{2}}\int \frac{ d^2 z}{2\pi} \sum_{i=1}^{n} 
\sum_{j=1}^{n} \left[ \left(  \frac{ \theta_i(\epsilon_q \epsilon_i)}{(z-z_i)^2}
\right) \left(\frac{({\bar{\epsilon}}_q k_j)}{{\bar{z}}-{\bar{z}}_j}  \right) +
\left( \frac{ {\bar{\theta}}_i({\bar{\epsilon}}_q {\bar{\epsilon}}_i)}{({\bar{z}}- 
{\bar{z}}_i)^2} \right)
\left(\frac{(\epsilon_q k_j)}{z-z_j} \right)  \right] \prod_{\ell=1}^{n} 
|z- z_{\ell}|^{\alpha' q k_{\ell}} 
\\
& \times  \Bigg\{ 
1
-  
\left(  
\frac{\sqrt{2\alpha'}}{2}   \right)   \sum_{k=1}^{n} \Big( 
\theta_k \frac{\epsilon_k q}{z-z_k} 
+
{\bar{\theta}}_k \frac{({\bar{\epsilon}}_k q)  }{{\bar{z}}-{\bar{z}}_k} 
\Big) 
 + \frac{1}{2}\left( \frac{\alpha'}{2}\right )
\\
& 
\times \Bigg[
\left( \sum_{h=1}^{n}   \theta_h  \frac{(\epsilon_h q)  }{z-z_h}  \right)^2
+
\left( \sum_{h=1}^{n}   {\bar{\theta}}_h \frac{({\bar{\epsilon}}_h q)
  }{{\bar{z}}-{\bar{z}}_h}   \right)^2
+  
2\left( \sum_{h=1}^{n}   \theta_h  \frac{(\epsilon_h q)  }{z-z_h}  \right) \left( \sum_{h=1}^{n}   {\bar{\theta}}_h \frac{({\bar{\epsilon}}_h q)  }{{\bar{z}}-{\bar{z}}_h}   \right)
\Bigg]  \Bigg\}\, .
  \label{S_3}
  }
These terms provide all contributions to the order $q^1$.  
They can be  further split in $S_i^{(a)}$, a=0,1,2, with the 
index $a$  labelling  the order of expansion in $q$ of the integrand modulo the factor $|z-z_l|^{\aqk{l}}$, which has to be integrated. The integrals involved are all of the form:
\ea{
I_{i_1 i_2 \ldots}^{j_1 j_2 \ldots} =
\int \frac{d^2 z}{2 \pi} \frac{\prod_{l = 1}^n |z-z_l|^{\alpha' k_l q}}{
(z-z_{i_1})(z-z_{i_2}) \cdots (\bar{z}-\bar{z}_{j_1}) (\bar{z}-\bar{z}_{j_2}) \cdots } \ .
\label{GeneralIntegralintext}
}
Each of the integrals involved has to be 
computed through order $q^{1-a}$, which we denote by ${I^{(1-a)}}_{i_1 i_2 \ldots}^{j_1 j_2 \ldots}$. Using this notation, each term can be compactly expressed as:
\seal{Sl1}{
S_1^{(0)} &= \frac{\alpha'}{2} \sum_{i=1}^n \left [
(\epsilon_qk_i)(\bar{\epsilon}_qk_i ) {I^{(1)}}_i^i +
\sum_{j\neq i}^n (\epsilon_qk_i)(\bar{\epsilon}_qk_j ) {I^{(1)}}_i^j \right ],
\label{S1su0giu}
\\
S_1^{(1)} &= -\left(\frac{\alpha'}{2}\right)^{\frac{3}{2}} \sum_{i,j,l=1}^n 
(\epsilon_qk_i)(\bar{\epsilon}_qk_j )(\theta_l\epsilon_lq){I^{(0)}}^j_{il}+\text{c.c.} \, ,
\\
S_1^{(2)} &=\frac{1}{2} \left(\frac{\alpha'}{2}\right)^2\sum_{i,j,l=1}^n
(\epsilon_qk_i)(\bar{\epsilon}_qk_j) (\theta_l\epsilon_lq)\left[ \sum_{m\neq l}^n 
(\theta_m\epsilon_mq) {I^{(-1)}}^j_{ilm} +
\sum_{m=1}^n (\bar{\theta}_m\bar{\epsilon}_mq)
 {I^{(-1)}}^{jm}_{il}\right] 
+\text{c.c.} \, ,
}
with $\text{c.c.}$  denoting the complex conjugate of the expressions. 
Similarly:
\seal{Sl2}{
S_2^{(0)} =&\, \sum_{i,j=1}^n(\epsilon_q\theta_i\epsilon_i)(\bar{\epsilon}_q\bar{\theta}_j\bar{\epsilon}_j){I^{(1)}}^{jj}_{ii} \, , \\
S_2^{(1)} =&\,  -\sqrt{\frac{\alpha'}{2}} \sum_{i,j=1}^n \sum_{l\neq i=1}^n (\epsilon_q\theta_i\epsilon_i)(\bar{\epsilon}_q\bar{\theta}_j\bar{\epsilon}_j)(\theta_l\epsilon_lq){I^{(0)}}^{jj}_{iil}+\text{c.c.}\, ,\\
S_2^{(2)}=&\, \frac{\alpha'}{4}\sum_{i,j=1}^n (\epsilon_q\theta_i\epsilon_i)(\bar{\epsilon}_q\bar{\theta}_j\bar{\epsilon}_j)\left[ \sum_{l\neq m\neq i=1}^n (\theta_l\epsilon_lq)(\theta_m\epsilon_mq) {I^{(-1)}}^{jj}_{iilm} \right.
\nonumber \\
&\left. +\sum_{l\neq i=1}^n\sum_{m\neq j=1}^n(\theta_l\epsilon_lq)(\bar{\theta}_m\bar{\epsilon}_mq) {I^{(-1)}}^{jjm}_{iil}\right]+\text{c.c.} \, ,
}
where we note that according to Eq.~(\ref{A88}) the first part of  $S_2^{(2)}$ involving $I_{iilm}^{jj}$ does not contribute to the order $q$ for any $i,j$ and $l\neq m\neq i$.
Finally:
\seal{Sl3}{
S_3^{(0)}=&\, \sqrt{\frac{\alpha'}{2}}\sum_{i,j=1}^n (\epsilon_q\theta_i\epsilon_i)(\bar{\epsilon}_qk_j){I^{(1)}}^j_{ii}+\text{c.c.} \,  , \\
S_3^{(1)}=&\,-\frac{\alpha'}{2} \sum_{i,j=1}^n (\epsilon_q\theta_i\epsilon_i)(\bar{\epsilon}_qk_j)\left[\sum_{l\neq i=1}^n (\theta_l\epsilon_lq){I^{(0)}}^j_{iil} +\sum_{l =1}^n(\bar{\theta}_j\bar{\epsilon}_jq){I^{(0)}}^{jl}_{ii}\right]+\text{c.c.}\, , \\
S_3^{(2)} =&\, \frac{1}{2}\left(\frac{\alpha'}{2}\right)^{\frac{3}{2}}\sum_{i,j=1}^n (\epsilon_q\theta_i\epsilon_i)(\bar{\epsilon}_qk_j) \left[\sum_{l\neq m\neq i=1}^n (\theta_l\epsilon_lq)(\theta_m\epsilon_mq){I^{(-1)}}^j_{iilm}\right.\nonumber\\
& \left.
+
\sum_{l \neq m = 1} (\tbeq{l})(\tbeq{m} ){I^{(-1)}}_{ii}^{jlm} 
+
\sum_{l\neq i=1}^n\sum_{m\neq j=1}^n (\theta_l\epsilon_lq)(\bar{\theta}_m\bar{\epsilon}_mq) {I^{(-1)}}^{jm}_{iil}\right]+\text{c.c.} \, ,
}
where we note that by inspection of Eqs.~\eqref{Iiimii}-\eqref{Iiimjn} the second part of $S_3^{(2)}$ involving $I_{ii}^{jlm}$ does not contribute to the order $q$ for any $j, l$, and $m\neq l$.

As a word of 
warning, notice that the definitions of $S_i$ are not the same as 
in Ref.~\cite{DiVecchia:2015oba}, but one can identify $S_1^{(0)}$, $S_1^{(1)}$, $S_2^{(0)} + S_2^{(1)}$, 
and $S_3^{(0)} + S_3^{(1)}$, with respectively $S_1, S_2, S_4$ and $S_3$ 
of Ref.~\cite{DiVecchia:2015oba}. 
In App.~\ref{Results} we provide the computational details as well as the explicit results for all the integrals involved. In particular, we show in the appendix 
that all the integrals are linear combinations  of a subset of six of them.
The coefficients of these linear combinations are  complex functions  with poles when two Koba-Nielsen variables coincide.
We now report the results. 

The first term of $S_1$, i.e. $S_1^{(0)}$,
is the part equivalent to the amplitude of a soft massless string scattering 
on $n$ tachyons, and this was already computed to the order $q^1$ 
in Ref.~\cite{DiVecchia:2015oba}, reading:
\ea{
S_1^{(0)}=&\, 
\epsilon_{q}^{S\mu \nu}
\Bigg \{ \sum_{i=1}^n k_{i\mu}k_{i\nu}
 \Bigg[\frac{(\alpha')^2}{2} \sum_{j \neq i} (k_j q) \log^2 |z_i - z_j|  
 \nonumber \\
&
 + \frac{1}{k_i q}  
\left( 1 +\alpha'  \sum_{j \neq i} (k_j q) \log |z_i - z_j|  
 + \frac{(\alpha')^2}{2} 
\sum_{j \neq i} \sum_{k \neq i} (k_j q) (k_k q)  \log|z_i -z_j| \log |z_i - z_k| \right) \Bigg]
\nonumber \\
&
-
\alpha'\sum_{i \neq j}^n k_{i\mu}k_{j\nu}
 \Bigg[ \log|z_i-z_j|-\frac{\alpha' }{2}\sum_{m\neq i,j}(q k_m)\log|{z}_m-{z}_j|\log|z_i-z_m| \nonumber\\
&+ \frac{\alpha' }{2}\sum_{m\neq j} (q k_m)\log|{z}_m-{z}_j|\log|z_i-z_j|
+
\frac{\alpha' }{2}\sum_{m\neq i}(q k_m)\log|{z}_i-{z}_j|\log|z_i-z_m| \Bigg] \Bigg \}
\nonumber \\
&
+
 \epsilon_{q}^{B\mu\nu} \sum_{i\neq j\neq m}^n k_{i\mu}k_{j\nu}
\left (\frac{\alpha'}{2}\right )^2  (q k_m) 
\Bigg [{\rm Li}_2\left( \frac{\bar{z}_i-\bar{z}_m}{\bar{z}_i-\bar{z}_j}\right)-{\rm Li}_2\left(\frac{z_i-z_m}{z_i-z_j }\right)
\nonumber\\
& 
+
\log\frac{|z_i-z_j|}{|z_i-z_m|}\log\left(\frac{z_m-z_j}{\bar{z}_m-\bar{z}_j}
\frac{\bar{z}_i-\bar{z}_j}{z_i-z_j}\right) \Bigg ]
+ \Ord(q^2)
\ ,
\label{S10}
}
where
\ea{
\epsilon_{q}^{S\mu \nu} = \frac{ \epsilon_q^\mu \bar{\epsilon}_q^\nu + 
 \epsilon_q^\nu \bar{\epsilon}_q^\mu}{2} \ , \ 
\epsilon_{q }^{B\mu \nu} = \frac{ \epsilon_q^\mu \bar{\epsilon}_q^\nu - 
 \epsilon_q^\nu \bar{\epsilon}_q^\mu}{2} \ .
\label{poldecomposition}
}
For the next terms, only the parts up to order $q^0$ were derived 
previously. For completeness we express the full result, together with 
the new terms of order $q$:
\ea{
S_1^{(1)} = &
-  
 \epsilon_{q \mu} \bar{\epsilon}_{q \nu}
 \sqrt{\frac{\alpha'}{2}}
 \sum_{i \neq j}
 \Bigg [ 
\frac{\theta_i \epsilon_i q}{z_i - z_j}
\Bigg ( 
\frac{ k_j^\mu k_i^\nu}{ q k_i} - \frac{k_j^\mu k_j^\nu}{ q k_j} 
+\a' (k_j^\mu k_i^\nu -k_j^\mu k_j^\nu )\log | z_i - z_j |
\nonumber \\
&
+
{\frac{\a'}{2} k_i^\mu k_i^\nu\frac{ q k_j }{ q k_i}}
 -\frac{\a'}{2} k_i^\mu k_j^\nu 
+
\a' k_j^\mu k_i^\nu \sum_{l\neq i} \frac{ q k_l}{q k_i} \log | z_i - z_l |
-
\a' k_j^\mu k_j^\nu \sum_{l\neq j} \frac{ q k_l}{q k_j} \log | z_j - z_l |
\Bigg )
\nonumber \\
&
+ \a' k_i^\mu k_j^\nu \sum_{l\neq ij} \theta_l \epsilon_l q \frac{\log |z_j-z_l| - \log |z_i-z_j|}{{z}_i - {z}_l}
\Bigg ] + \text{c.c.} \, ,
}
\ea{
S_1^{(2)} = &
 \epsilon_{q \mu} \bar{\epsilon}_{q \nu}
 \frac{\a'}{2} \sum_{i\neq j}^n
 \Bigg [
 \frac{k_j^\mu k_i^\nu - k_i^\mu k_i^\nu}{qk_i} \frac{(\theta_i \epsilon_i q )(\theta_j \epsilon_j q)}{(z_i-z_j)^2}
  + \frac{k_i^\mu k_j^\nu}{2} \sum_{l\neq i,j}^n \frac{(\teq{l})(\tbeq{l})}{qk_l (\zz{i}{l})(\zbzb{j}{l})}
   \qquad \qquad 
 \nonumber \\
& +  
 \left (k_i^\mu k_i^\nu (\tbeq{j}) + k_i^\mu k_j^\nu (\tbeq{i}) \right )
 \frac{ \teq{j}}{2|z_i-z_j|^2}\left(\frac{ 1}{\qk{i}} + \frac{1}{\qk{j}} \right ) 
 \nonumber \\
 &  + \frac{\frac{k_i^\mu k_i^\nu}{2} (\theta_j \epsilon_j q)
+ k_j^\mu k_i^\nu (\theta_i \epsilon_i q)}{qk_i (\zz{i}{j})} \sum_{l\neq i,j} 
\left ( \frac{\teq{l}}{(z_i-z_l)} + \frac{\tbeq{l}}{(\zbzb{i}{l})}\right )
 \Bigg] + \text{c.c.} \, ,
 \label{l11}}
\seal{l12}{
S_2^{(0)} =&\,
 \epsilon_{q \mu} \bar{\epsilon}_{q \nu} \frac{\alpha'}{2}
 \sum_{i \neq j}^n 
 \frac{(\theta_i \epsilon_i^\mu)}{|\zz{i}{j}|^2}
\Bigg \{(\bar{\theta}_i \bar{\epsilon}_i^\nu)qk_j
\Bigg (1
+\frac{1}{2}\sum_{l\neq i}
 \frac{\qk{l}}{\qk{i}}
\left [
\frac{\zbzb{i}{j}}{\zbzb{i}{l}}+
\frac{\zz{i}{j}}{\zz{i}{l}} \right ]
\Bigg )
\nonumber \\
&
-(\bar{\theta}_j \bar{\epsilon}_j^\nu) 
 \sum_{l \neq i,j} qk_l
\frac{(\zbzb{i}{l})(\zz{j}{l})}{(\zz{i}{l})(\zbzb{j}{l})}
\Bigg\} \, , \\[5mm]
S_2^{(1)} =&\,
 \epsilon_{q \mu} \bar{\epsilon}_{q \nu}
\sqrt{\frac{\a'}{2}}\sum_{i\neq j}^n \Bigg[
 \frac{
\left (\temu{i}  (\teq{j})
-
\temu{j}  (\teq{i}) \right )(\tbenu{i} )
}{|\zz{i}{j}|^2 (\zz{i}{j})} 
\left(1+  \sum_{l\neq i} \frac{\qk{l}}{\qk{i}}
\frac{\zbzb{i}{j}}{\zbzb{i}{l}} \right )
\nonumber \\
&+ 
\frac{\temu{i}(\tbenu{j})}{\zbzb{i}{j}}
\sum_{l\neq i,j}^n
 \frac{(\theta_l \epsilon_l q)(\zbzb{i}{l})}{(\zz{i}{l})^2(\zbzb{j}{l}) }\Bigg ]
+ \text{c.c.} \, ,\\[5mm]
S_2^{(2)} =&\,
\epsilon_{q\mu} \bar{\epsilon}_{q\nu}
\sum_{i \neq j }\frac{(\theta_i \epsilon_i^\mu)(\theta_j \epsilon_j q)}{2(\zz{i}{j})^2}
\Bigg [
\frac{1}{\qk{i}}\sum_{l\neq i}\Bigg(
\frac{ (\bar{\theta}_i \bar{\epsilon}_i^\nu)
(\bar{\theta}_l \bar{\epsilon}_l q)- (\bar{\theta}_l\bar{\epsilon}_l^\nu)
(\bar{\theta}_i\bar{\epsilon}_i q)}
{ (\zbzb{i}{l})^2}
 \nonumber \\
&
+ \frac{1}{\qk{j}}\sum_{l\neq j}
 \frac{ (\bar{\theta}_l\bar{\epsilon}_l^\nu)
(\bar{\theta}_j\bar{\epsilon}_j q)-(\bar{\theta}_j\bar{\epsilon}_j^\nu)
(\bar{\theta}_l\bar{\epsilon}_l q)}{(\zbzb{l}{j})^2}
\Bigg) \Bigg] + \text{c.c.} \, ,
}
\seal{I13}{
S_3^{(0)} = &\,
 \epsilon_{q \mu} \bar{\epsilon}_{q \nu}
\sqrt{\frac{\alpha'}{2}}
 \sum_{i\neq j}^n 
\a'q_\rho \Bigg [
 \frac{ k_i^\nu k_j^\rho - k_j^\nu k_i^\rho}{\zz{i}{j}}\temu{i}
\left( \frac{1}{\aqk{i}} +\frac{1}{2} 
\right)
\nonumber \\
&
+ \sum_{l\neq i} 
\frac{k_i^\nu k_l^\rho-k_l^\nu k_i^\rho}{\zz{i}{j}} \left (\temu{j} + \temu{i} \frac{\qk{j}}{\qk{i}} \right )
\LN{i}{l}
\Bigg ]+\text{c.c.} \, ,
\\[5mm]
S_3^{(1)} = &\,
 \epsilon_{q \mu} \bar{\epsilon}_{q \nu}
 \sum_{i\neq j}^n  \temu{i}\Bigg\{
\frac{ (\theta_j \epsilon_j q)}{(z_i - z_j)^2}
 \Bigg [
 \frac{k_{i}^{\nu}}{k_i q} - \frac{k_{j}^{\nu}}{k_j q} 
  -\a' q_\rho \sum_{l\neq j} 
 \frac{k_j^\nu k_l^\rho - k_l^\nu k_j^\rho}{\qk{j}} \LN{j}{l}
  \nonumber \\
 &
 -\a' q_\rho \sum_{l\neq i}
 \frac{k_i^\nu k_l^\rho - k_l^\nu k_i^\rho}{\qk{i}}  
 \left (
 \frac{1}{2} \frac{\zz{i}{j}}{\zz{i}{l}} - 
 \LN{i}{l} 
 \right) 
 \Bigg] 
   - \frac{\alpha'}{2}
\sum_{l\neq i,j}
\frac{k_j^\nu \tbeq{l} + k_l^\nu \tbeq{j}}{(\zz{i}{l})(\zbzb{j}{l})}
 \nonumber \\
 &
  - \frac{\alpha'}{2}
\frac{k_i^\nu \bar{\theta}_j \bar{\epsilon}_j q
+  
k_j^\nu \bar{\theta}_i \bar{\epsilon}_i q
}{|z_i-z_j|^2} 
\left (1 + \sum_{l\neq i} \frac{\qk{l}}{\qk{i}} \frac{\zz{i}{j}}{\zz{i}{l}} \right )
 \Bigg\} +\text{c.c.} \, ,
 \\[5mm]
 S_3^{(2)} = &\,
  \epsilon_{q \mu} \bar{\epsilon}_{q \nu}
\sqrt{\frac{\alpha'}{2}} \sum_{i\neq j}^n 
\Bigg [ \sum_{l\neq i,j}^n 
\frac{ \temu{i}
(\theta_j \epsilon_j q) (\theta_l \epsilon_l q)}{(\zz{j}{l})(\zz{i}{j})^2}
\left (\frac{ k_j^\nu}{\qk{j}} -\frac{k_i^\nu}{\qk{i}} \right )
  \nonumber \\
&
+
\frac{ \temu{j}(\teq{i}) -\temu{i}(\teq{j})}{\qk{i}} 
\sum_{l\neq i} 
\frac{\left (k_i^\nu \tbeq{l} + k_l^\nu \tbeq{i} \right )}{(\zz{i}{j})^2 (\zbzb{i}{l})} 
 \Bigg ] + {\rm c.c.} \, .
}
As a nontrivial consistency check, it is possible to show that the full 
expression \mbox{$S_1 + S_2 + S_3$} obeys gauge invariance, meaning 
that it vanishes identically by the replacement $\epsilon_{q\mu} \to q_\mu$ 
and $\bar{\epsilon}_{q\nu} \to q_\nu$. Actually, the identity is stronger, since 
the full expression vanishes by replacing only $\epsilon_{q\mu} \to q_\mu$ 
\emph{or} $\bar{\epsilon}_{q\nu} \to q_\nu$, which can be explicitly checked from the above expression. In other words, 
\ea{
q_\mu M_{n+1}^{\mu \nu} = q_\nu M_{n+1}^{\mu \nu} = 0 \, ,
\label{gaugeinvariance}
}
where $M_{n+1}^{\mu \nu}$ is the stripped soft amplitude with respect to 
the polarization of the soft particle.

We want to find a gauge invariant operator that, when acting on $M_n$ reproduces the above results, i.e.
\ea{
M_{n+1}(q; k_i) = M_n(k_i) \ast S(q,k_i) 
&= \kappa_D \left ( \hat{S}_q^{(-1)}+\hat{S}_q^{(0)}+\hat{S}_q^{(1)} \right ) M_n(k_i) + \Ord(q^2) \, ,
\label{softexpansion}
}
where the superscript of each $\hat{S}_q^{(m)}$ indicates the order $m$ in $q$ of each term.
In Ref.~\cite{DiVecchia:2015oba} we showed that the leading and 
subleading terms, symmetric in the polarization indices $\mu, \nu$, are 
generated by exactly the same soft-operators that one can infer using just gauge-invariance of the amplitude, which read:
\ea{
\hat{S}_q^{(-1)} &= \epsilon_{\mu \nu}^S \sum_{i=1}^n \frac{k_i^\mu k_i^\nu}{k_i \cdot q}
\ , \quad
\hat{S}_q^{(0)} =  \epsilon_{\mu \nu}^S\left ( - \frac{i q_\rho}{2} \right )\sum_{i=1}^n \frac{k_i^\mu J_i^{\nu \rho} + k_i^\nu J_i^{\mu \rho}}{k_i \cdot q} \, ,
}
where
\begin{eqnarray}
J_i^{\mu \nu} = L_i^{\mu \nu} + 
\mathcal{S}_i^{\mu \nu} \, , \quad
\mathcal{S}_i^{\mu \nu} = S_i^{\mu \nu} 
+ {\bar{S}}^{\mu \nu}_i 
\ ,
\label{JLS}
\end{eqnarray}
\ea{
L_i^{\mu\nu} =i\left( k_i^\mu\frac{\partial }{\partial k_{i\nu}} -k_i^\nu\frac{\partial }{\partial k_{i\mu}}\right)  , \ 
S_i^{\mu\nu}=i\left( \epsilon_i^\mu\frac{\partial }{\partial \epsilon_{i\nu}} -\epsilon_i^\nu\frac{\partial }{\partial \epsilon_{i\mu}}\right)  , \  
{\bar{S}}^{\mu\nu}_i=i\left( {\bar{\epsilon}}_i^\mu\frac{\partial }{\partial {\bar{\epsilon}}_{i\nu}} -{\bar{\epsilon}}_i^\nu\frac{\partial }{\partial {\bar{\epsilon}}_{i\mu}}\right)  \, .
\label{LandS1}
}

The new result here is that the subsubleading terms, symmetric in the 
polarization indices $\mu, \nu$, are uniquely generated by the following 
soft operator, which can be explicitly checked:
\ea{
S_q^{(1)} = &- \frac{\epsilon_{\mu \nu}^S}{2} \sum_{i=1}^n
\left [
\frac{q_\rho J_i^{\mu \rho} q_\sigma J_i^{\nu \sigma}}{k_i \cdot q}
+ \left (
\frac{k_i^\mu q^\nu}{k_i \cdot q} q^\sigma + q^\mu \eta^{\nu \sigma}- \eta^{\mu \nu} q^\sigma \right )
\frac{\partial}{\partial k_{i}^\sigma} \right .
\nonumber \\
&-\left (
\frac{q_\rho q_\sigma \eta_{\mu \nu} - q_\sigma q_\nu \eta_{\rho \mu} - q_\rho q_\mu \eta_{\sigma \nu}}{ k_i\cdot q}
\right )  \left (\epsilon_{i}^\rho \frac{\partial}{\partial \epsilon_{i\sigma}} + \bar{\epsilon}_{i}^\rho \frac{\partial}{\partial \bar{\epsilon}_{i\sigma}} \right )
\nonumber \\
&\left .  - \alpha' \left (q_\sigma k_{i\nu} \eta_{\rho \mu}+q_\rho k_{i\mu }\eta_{\sigma \nu} - \eta_{\rho\mu}\eta_{\sigma \nu} (k_i \cdot q) - q_\rho q_\sigma \frac{k_{i\mu}k_{i\nu}}{k_i \cdot q} \right )
\left (\epsilon_{i}^\rho \frac{\partial}{\partial \epsilon_{i\sigma}} + \bar{\epsilon}_{i}^\rho \frac{\partial}{\partial \bar{\epsilon}_{i\sigma}} \right )
  \right ] .
  \label{generalsubsub}
}
It is thus useful to also define:
\ea{
\Pi_i^{\rho \sigma} = 
\epsilon_i^\rho\frac{\partial }{\partial \epsilon_{i\sigma}} +
{\bar{\epsilon}}_i^\rho\frac{\partial }{\partial {\bar{\epsilon}}_{i\sigma}} \, .
\label{sumpoloperator}
}
Notice that only the symmetric combination $\Pi_i^{\{\rho, \sigma\}} = 
\frac{\Pi_i^{\rho \sigma} + \Pi_i^{\sigma \rho}}{2}$ survives the contractions 
in Eq.~\eqref{generalsubsub}, since the contraction of $\mu$ and $\nu$ is symmetric.

The terms in the first two lines of Eq.~\eqref{generalsubsub}, which are finite in the field theory limit, exactly match the soft theorem derived in Ref.~\cite{DiVecchia:2015jaq} using just on-shell gauge invariance of tree-level gravity amplitudes. The terms in the last line can thus be seen as the string corrections to the field theory soft theorem. 
Notice that each parenthesis is independently gauge invariant. 
Notice also that in the field theory limit, if the soft particle is a graviton, only the first term is nonzero, since $\epsilon_{\mu \nu}^{\rm graviton}$ is traceless.
The extra terms in the first line were found already in Ref.~\cite{DiVecchia:2015oba} for the case, where the $n$ external states were tachyons.

In Ref.~\cite{DiVecchia:2015oba}, we also found a soft theorem for the 
antisymmetric part at the subleading order, corresponding to a soft 
Kalb-Ramond field. At this point, however, it is not clear how the antisymmetric 
part of our subsubleading explicit results could also be expressed as a soft 
theorem, since at this order dilogarithmic terms appear in Eq.~\eqref{S10}. 
We thus leave the analysis of the antisymmetric part for a possible future study.

In the next section we specify the subsubleading operator to the case of a soft
dilaton and a soft graviton and we give a physical interpretation of the various
terms that appear.

\section{Soft gravitons and dilatons}
\label{softdilasoftgravi}

Specifying our main result Eq.~\eqref{generalsubsub} to the
 cases where the soft particle is either a graviton or a dilaton,
we may first simplify the general expression by imposing the 
transversality condition $\epsilon_{\mu \nu}^S q^\mu = \epsilon_{\mu \nu}^S q^\nu = 0$, 
leading to:
\seq{
S_q^{(1)} = &- \frac{\epsilon_{\mu \nu}^S}{2} \sum_{i=1}^n
\left [
\frac{q_\rho J_i^{\mu \rho} q_\sigma J_i^{\nu \sigma}}{k_i \cdot q}
- \frac{\eta^{\mu \nu}   q_\rho q_\sigma}{k_i \cdot q} \left (
k_i^\rho
\frac{\partial}{\partial k_{i\sigma}} 
+
\Pi_i^{\{\rho, \sigma\}}
\right )
\right .
\\
&\left .  - \alpha' \left (q_\sigma k_{i\nu} \eta_{\rho \mu}+q_\rho k_{i\mu }\eta_{\sigma \nu} - \eta_{\rho\mu}\eta_{\sigma \nu} (k_i \cdot q) - q_\rho q_\sigma \frac{k_{i\mu}k_{i\nu}}{k_i \cdot q} \right )
\Pi_i^{\{\rho, \sigma\}}
  \right ] .
  \label{subsub}
}
Considering the soft particle to be a graviton, tracelessness of its polarization gives:
\seq{
S_{{\rm graviton}, q}^{(1)} = &- \frac{\epsilon_{\mu \nu}^{\rm graviton}}{2} \sum_{i=1}^n
\left [
\frac{q_\rho J_i^{\mu \rho} q_\sigma J_i^{\nu \sigma}}{k_i \cdot q}
\right .
 \\
&\left .  - \alpha' \left (q_\sigma k_{i\nu} \eta_{\rho \mu}+q_\rho k_{i\mu }
\eta_{\sigma \nu} - \eta_{\rho\mu}\eta_{\sigma \nu} (k_i \cdot q) - 
q_\rho q_\sigma \frac{k_{i\mu}k_{i\nu}}{k_i \cdot q} \right )
\Pi_i^{\{\rho, \sigma\}}
  \right ] .
  \label{softgraviton}
}
The first term reproduces the subsubleading soft theorem of gravitons.
The second line are the string corrections to the field theory result. We can reduce the derivatives with respect to $\epsilon_i$ and $\bar{\epsilon}_i$ in $\Pi_i$ by acting on the 
 $n$-point amplitude with the polarization vectors stripped off, i.e.
\ea{
M_n(k_i , \epsilon_i, \bar{\epsilon}_i) = \epsilon_1^{\mu_1}\bar{\epsilon}_1^{\nu_1} \cdots \epsilon_n^{\mu_n}\bar{\epsilon}_n^{\nu_n} M_{n,(\mu_1,\nu_1), \ldots, (\mu_n,\nu_n)} (k_i) \, .
\label{strippedMn}
}
Then we can express:
\seq{
\left (\epsilon_{i}^\rho \frac{\partial}{\partial \epsilon_{i\sigma}} 
+ \bar{\epsilon}_{i}^\rho \frac{\partial}{\partial \bar{\epsilon}_{i\sigma}} 
\right ) M_n &= 
\left ( \eta^{\sigma \mu_i} \epsilon_i^\rho \bar{\epsilon}_i^{\nu_i} + \eta^{\sigma \nu_i} \epsilon_i^{\mu_i} \bar{\epsilon}_i^{\rho} \right )
M_{n, (\mu_i \nu_i)} \\
&=2
\eta^{\sigma \mu_i} \left ( \epsilon_i^{\{\rho,} \bar{\epsilon}_i^{\nu_i \}} M_{n,\{\mu_i, \nu_i\}}
+  \epsilon_i^{[\rho,} \bar{\epsilon}_i^{\nu_i ]} M_{n,[\mu_i, \nu_i]} \right ) \, ,
\label{gravitonstringpole}
}
where  in the second line we decomposed $M_n$ into its symmetric and 
antisymmetric parts, as in Eq.~(\ref{poldecomposition}), showing that string 
corrections can exist for external states $i$ being polarized both  symmetrically 
(gravitons and dilatons) and antisymmetrically (Kalb-Ramond).
We will comment further on these new string-theory terms in the next section.

Projecting instead the soft leg onto the dilaton, using $\epsilon_{\mu \nu}^d = 
(\eta_{\mu \nu} - q_\mu \bar{q}_\nu - q_\nu \bar{q}_\mu)/\sqrt{D-2}$, 
with $q\cdot \bar{q}= 1$ and $\bar{q}^2=q^2 = 0$, we get:
\ea{
S_{{\rm dilaton}, q}^{(1)} =  \frac{1}{2\sqrt{D-2}}  \sum_{i=1}^n
&\left[ q^\rho {\hat{K}}_{i \rho} 
 +
\frac{q^\rho q^\sigma}{k_i q}
\left( 
\mathcal{S}_{i, \rho \mu}
\eta^{\mu \nu} 
\mathcal{S}_{i \nu \sigma}
+ D 
\Pi_{i,\{\rho, \sigma\}}
\right) 
\right .
\nonumber \\
& \left .
-\alpha'(k_i \cdot q) \left (
 \epsilon_i \cdot \frac{\partial}{\partial \epsilon_i} + \bar{\epsilon}_i \cdot \frac{\partial}{\partial \bar{\epsilon}_i} \right )
\right] \, ,
\label{subsubdilaton}
}
where both $k_i \cdot \epsilon_i = k_i \cdot \bar{\epsilon}_i = 0$ and 
gauge invariance, i.e. $k_i\cdot \frac{\partial}{\partial \epsilon_i}M_n = 
k_i^\mu M_{n, \mu} = 0$ and $k_i\cdot \frac{\partial}{\partial 
\bar{\epsilon}_i} M_n = k_i^\nu M_{n, \nu} = 0$, were used, and
 where we introduced the operator:
\begin{eqnarray}
&&  {\hat{K}}_{i\mu} = 2 \left[ \frac{1}{2} k_{i \mu} \frac{\partial^2}{\partial
k_{i\nu} \partial k_i^\nu} 
-k_{i}^{\rho} \frac{\partial^2}{\partial
k_i^\mu \partial k_{i}^{\rho}} 
+ i
\mathcal{S}_{i,\rho \mu} \frac{\partial}{\partial k_i^\rho}  \right] \, .
\label{hatDhatKmu}
\end{eqnarray}
Remarkably, this is exactly the generator of special conformal transformations acting on momentum space.
The string correction for the dilaton vanishes due to
momentum conservation, since the operator $\epsilon_i \cdot \frac{\partial}{\partial \epsilon_i}$ leaves $M_n$ invariant, yielding (correspondingly for the barred term)
\ea{
\frac{\alpha'}{2}\sum_{i=1}^n (k_i \cdot q) \epsilon_i \cdot \frac{\partial}{\partial \epsilon_i} M_n = \frac{\alpha'}{2}
\sum_{i=1}^n (k_i \cdot q) M_n = - \frac{\alpha'}{2} q^2 M_n = 0 \ .
}
In conclusion, we find that the subsubleading dilaton soft operator equals the 
field theory counterpart and reads:
\ea{
S_{{\rm dilaton}, q}^{(1)} = & \frac{1}{2\sqrt{D-2}}  \sum_{i=1}^n
\left[ q^\rho {\hat{K}}_{i \rho} 
 +
\frac{q^\rho q^\sigma}{k_i q}
\left(
\mathcal{S}_{i, \rho \mu}\eta^{\mu \nu} \mathcal{S}_{i \nu \sigma} + D
\Pi_{i,\{\rho, \sigma\}}
\right) 
\right] .
\label{finalsubsubdilaton}
}
The subsubleading dilaton soft theorem contains a finite piece, which can be 
fully expressed by the generator of a special conformal transformation
 and a singular piece only dependent on polarization derivatives.  

\begin{figure}[tb]
\includegraphics[width=0.2\textwidth]{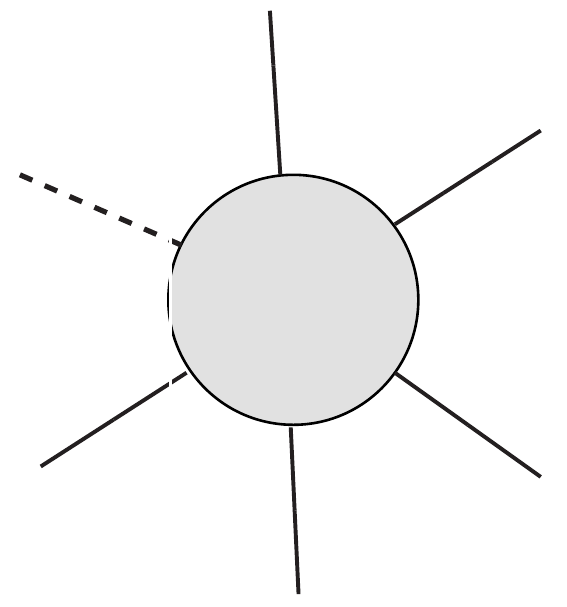}
\hspace{3mm}
\includegraphics[width=0.22\textwidth]{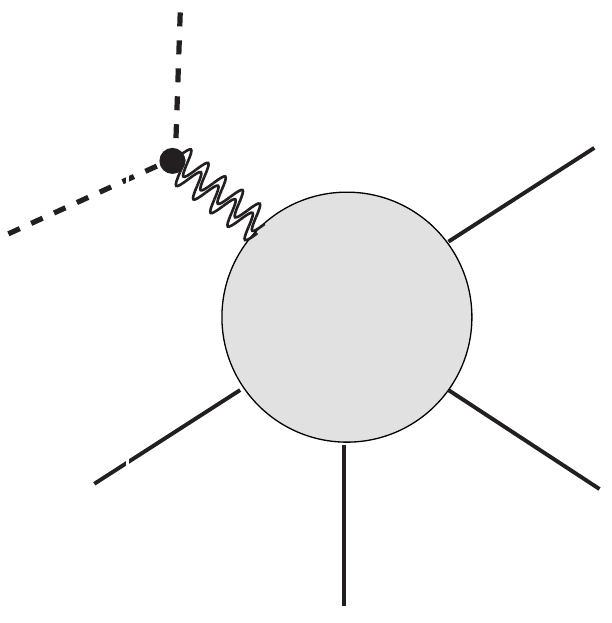}
\hspace{3mm}
\includegraphics[width=0.22\textwidth]{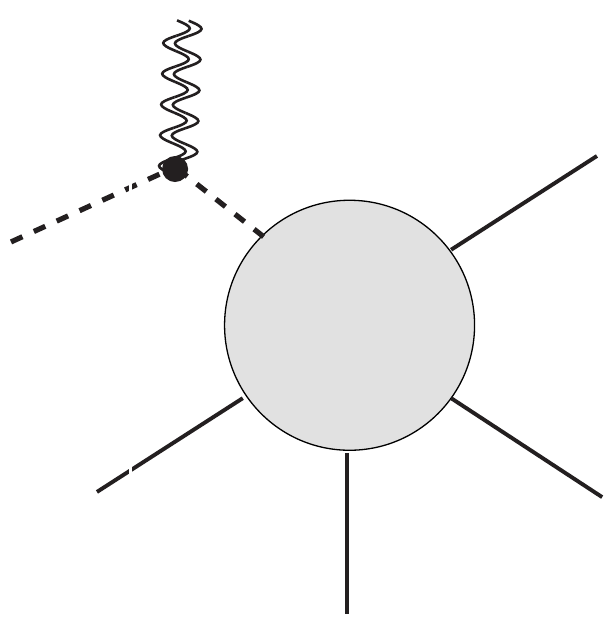}
\hspace{3mm}
\includegraphics[width=0.22\textwidth]{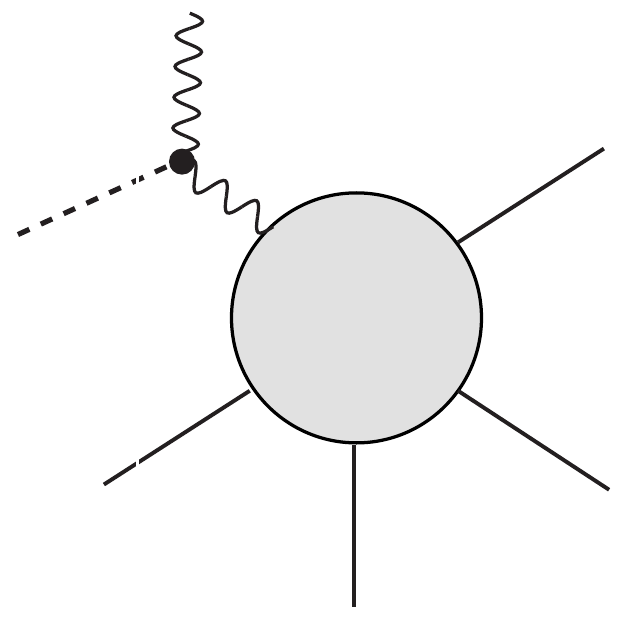}
\caption{Soft dilaton (dashed line) scattering on other massless states (solid lines), and the only three types of exchange diagrams appearing in field theory, involving i) another external dilaton and an internal graviton (double-wavy line), ii) an external graviton and an internal dilaton, iii) an external and an internal Kalb-Ramond field (wavy line). 
}
\label{fieldtheoryfactorization}
\end{figure}

We may use the polarization-stripped form of $M_n$ 
in Eq.~\eqref{strippedMn} to understand the singular terms, which then after some 
simplification read (suppressing for brevity the factor $1/\sqrt{D-2}$):
\seq{
&\sum_{i=1}^n \frac{q^\rho q^\sigma}{2 k_i q}
\left(
\mathcal{S}_{i, \rho \mu}\eta^{\mu \nu} \mathcal{S}_{i \nu \sigma} + D 
\Pi_{i,\{\rho, \sigma\}}
\right) M_n 
 \\
&
=
\sum_{i=1}^n \frac{1}{ k_i q} \Big (
q_{\mu_i}q_{\nu_i} (\epsilon_i \cdot \bar{\epsilon}_i) + \eta_{\mu_i \nu_i} (q \cdot \epsilon_i)(q \cdot \bar{\epsilon}_i) 
 \\
& \hspace{15mm} + (q \cdot \epsilon_i) (q_{\mu_i} \bar{\epsilon}_{\nu_i} - q_{\nu_i} \bar{\epsilon}_{\mu_i} )- 
(q \cdot \bar{\epsilon}_i) (q_{\mu_i} {\epsilon}_{\nu_i} - q_{\nu_i} {\epsilon}_{\mu_i} )
\Big ) M_n^{(\mu_i, \nu_i)} \, .
}
The expression evidently separates into a symmetric and an antisymmetric part, 
which we can express using $M_n^{\mu_i \nu_i} = M_n^{\{\mu_i, \nu_i\}} + M_n^{[\mu_i, \nu_i]}$, reducing the previous expression to:
\ea{
&\sum_{i=1}^n  \left (\frac{q_{\mu_i}q_{\nu_i} \eta_{\alpha \beta} + q_\alpha q_\beta \eta_{\mu_i \nu_i} }{ k_i \cdot q} 
\right ) \epsilon_i^\alpha \bar{\epsilon}_i^\beta M_n^{\{\mu_i, \nu_i\}} +
\sum_{i=1}^n 2\left (\frac{q_\alpha q_{\mu_i} \eta_{\beta \nu_i} + q_\beta q_{\nu_i} \eta_{\alpha \mu_i}}{ k_i \cdot q} 
\right )\epsilon_i^\alpha \bar{\epsilon}_i^\beta  M_n^{[\mu_i, \nu_i]}
\nonumber \\
=&
\sum_{i=1}^n  \left (\frac{q_{\mu_i}q_{\nu_i} \eta_{\alpha \beta} + q_\alpha q_\beta \eta_{\mu_i \nu_i} }{ k_i \cdot q} 
\right ) \epsilon_i^{S \alpha,\beta} M_n^{\{\mu_i, \nu_i\}} +
\sum_{i=1}^n 4\frac{ q_\alpha q_{\mu_i} \eta_{\beta \nu_i}}{ k_i \cdot q}\epsilon_i^{B \alpha,\beta}   M_n^{[\mu_i, \nu_i]} \, ,
}
where in the second line we also decomposed the polarization vectors 
as in Eq.~\eqref{poldecomposition}.
The form of these terms suggest that they are coming from factorizing 
exchange diagrams, where the soft dilaton is attached to an external leg 
through a three-point vertex, with the indices $\alpha, \beta$ being the 
polarization indices of the internal state.
In the field theory limit there are only two types of such vertices, one 
involving two dilatons derivatively coupled to a graviton and two Kalb-Ramond 
fields derivatively coupled to one dilaton, giving rise to the three types of 
factorizing diagrams shown in Fig.~\ref{fieldtheoryfactorization}.
Indeed, if we project the external leg $i$ on each of the three massless 
states, the expression above reduces in each case to one nonzero term:
\seal{softdilatonpoles}{
&\text{For } \epsilon_i^{\alpha \beta} = \epsilon_g^{\alpha\beta}:
 \qquad
\frac{ q_\alpha q_\beta}{ k_i \cdot q} 
  \, \eta_{\mu_i \nu_i} \, \epsilon_g^{\alpha,\beta} M_n^{\{\mu_i, \nu_i\}} \, .
  \\
  &\text{For } \epsilon_i^{\alpha \beta} = \epsilon_d^{\alpha\beta}:
 \qquad
\frac{ q_{\mu_i} q_{\nu_i} }{ k_i \cdot q} 
\, \eta_{\alpha\beta}  \, \epsilon_d^{\alpha,\beta}  M_n^{\{\mu_i, \nu_i\}} \, .
  \\
  &\text{For } \epsilon_i^{\alpha \beta} = \epsilon_B^{\alpha\beta}:
 \qquad
4 \frac{ q_{\alpha} q_{\mu_i} }{ k_i \cdot q} 
\, \eta_{\beta \nu_i}  \, \epsilon_B^{\alpha,\beta}  M_n^{[\mu_i, \nu_i]} \, .
} 

It is worth noticing from Fig.~\ref{fieldtheoryfactorization} that in the cases 
where the $i$th particle is either a dilaton or a graviton, there are contributions 
to the dilaton soft theorem, where the $i$th particle in the lower point amplitude 
$M_n$ is changed to respectively a graviton or a dilaton. 
This means that the subsubleading term of the
soft dilaton amplitude separates into two factorized terms when the $i$th state is a dilaton or a graviton. Specifically, for either of the three possible cases we have:
\ea{
M_{n+1}( \phi(q); \phi(k_i), \ldots) &\sim 
\left (\hat{S}^{(0)} + q^\mu \hat{S}^{(1)}_{ \mu} \right) M_n (\phi(k_i), \ldots )
+q^\mu \hat{S}^{\phi g}_{ \mu} \,  M_n (g_{\alpha \beta}(k_i), \ldots ) 
+ \Ord(q^2) \, , \nonumber \\[2mm]
M_{n+1}( \phi(q); g_{\mu_i\nu_i}(k_i), \ldots) &\sim 
\left (\hat{S}^{(0)} + q^\mu \hat{S}^{(1)}_{ \mu} \right) M_n (g_{\mu_i\nu_i}
(k_i), \ldots )
+q^\mu \hat{S}^{g\phi}_{ \mu} \,  M_n (\phi(k_i), \ldots ) + \Ord(q^2)
\, , \nonumber\\[2mm]
M_{n+1}( \phi(q); B_{\mu_i\nu_i}(k_i), \ldots) &\sim 
\left (\hat{S}^{(0)} + q^\mu \hat{S}^{(1)}_{ \mu} \right) M_n (B_{\mu_i\nu_i}(k_i), \ldots )+ \Ord(q^2) \, ,
}
where $\phi$ denotes a dilaton,  $g_{\alpha \beta}$ denotes a graviton, and $B_{\alpha \beta}$ denotes a Kalb-Ramond field. The non-standard factorizing behavior of the first two cases was also noticed in Ref.~\cite{BianchiR2phi} in specific examples.

In conclusion, in this section we have provided a  physical interpretation
of the subsubleading terms that survive in the field theory limit. In the 
next section we extend our analysis to the terms with string
corrections. In particular, we show that they are completely determined 
from gauge invariance and from the string correction terms, appearing in
the bosonic string, in the three-point amplitude involving massless particles.

\section{String corrections from gauge invariance}
\label{gaugeinvariance1}
In this section we derive
the string corrections to the soft operator, found in the previous section from explicit calculations,
by considering the string corrections to the three-point amplitude of the bosonic string involving three massless closed string and exploiting on-shell gauge invariance of the amplitude.
Let us
consider the three-point amplitude of three massless bosonic 
strings with the given set of momenta and polarizations:
\ea{
(q, \epsilon_q^\mu, \bar{\epsilon}_q^\nu) \, , \quad 
(k_i, \epsilon_i^{\mu_i}, \bar{\epsilon}_i^{\nu_i})  \, , \quad
(-k_i-q, \epsilon_m^{\alpha}, \bar{\epsilon}_m^{\beta}) \, .
}
The polarization stripped three-point on-shell amplitude then reads:
\seq{
M_3^{\mu\nu ;\,\mu_i\nu_i; \,\alpha\beta}= 
&2 \kappa_D
\left( \eta^{\mu\mu_i} q^\alpha-\eta^{\mu\alpha} q^{\mu_i} +\eta^{\mu_i\alpha} k_i^\mu -\frac{\alpha'}{2} k_i^\mu q^{\mu_i} q^\mu\right) \\
&\times\left( \eta^{\nu\nu_i} q^\beta-\eta^{\nu\beta} q^{\nu_i} +\eta^{\nu_i\beta} k_i^\nu -\frac{\alpha'}{2} k_i^\nu q^{\nu_i} q^\nu\right) \, .
\label{threepoint}
}
Contracting this expression with particular polarization tensors yields the 
explicit expression for particular three-point amplitudes of massless strings. 
For instance, considering the case where one of the states is a dilaton, and 
contracting with the polarization tensor used in Eq.~\eqref{subsubdilaton}, we
get the following nonzero three-point amplitudes involving one dilaton:
\seq{
M_{ddg} = 2 \kappa_D \epsilon_g^{\alpha \beta} q_\alpha q_\beta & \, , \quad
M_{dBB} = \frac{2 \kappa_D}{\sqrt{D-2}} \, 4 \epsilon_B^{\mu_i \nu_i} 
\eta_{\mu_i \alpha}\epsilon_B^{\alpha \beta} q_{\nu_i} q_\beta \, , \\
M_{dgg} &= - \alpha'\frac{2 \kappa_D}{\sqrt{D-2}} \, \epsilon_g^{\mu_i \nu_i} 
\epsilon_g^{\alpha \beta} q_{\mu_i} q_{\nu_i} q_\alpha q_\beta \, .
\label{3couplings}
}
Notice that the dilaton-graviton-graviton amplitude vanishes in the field theory limit.
From these three-point amplitudes we can immediately write down the contributions from the factorizing exchange diagrams in Fig.~\ref{fieldtheoryfactorization} to the soft theorem, i.e.
\ea{
M_{n+1}^{\rm ex.}&(\phi(q), \phi(k_i), \ldots ) \nonumber \\
\sim& \
M_{ddg}\, \frac{1}{(k_i + q)^2}\, M_n(g_{\alpha \beta}(k_i+q), \ldots ) 
=\kappa_D\, \frac{q_\alpha q_\beta}{k_i \cdot q}\, \epsilon_g^{\alpha \beta}M_n(g_{\alpha \beta}(k_i+q), \ldots ) \, , \\[5mm]
M_{n+1}^{\rm ex.}&(\phi(q), g_{\mu_i\nu_i}(k_i), \ldots ) 
\nonumber \\
\sim&\
M_{ddg} \,\frac{1}{(k_i + q)^2} \, M_n(\phi(k_i+q), \ldots ) 
+ M_{dgg} \,\frac{1}{(k_i + q)^2} \, M_n(g_{\alpha \beta} (k_i+q), \ldots ) 
\nonumber \\[2mm]
=&\  \kappa_D \, \frac{q_{\mu_i} q_{\nu_i}}{k_i \cdot q}\,  \epsilon_g^{\mu_i \nu_i} M_n(\phi(k_i+q), \ldots ) 
 -  \frac{\alpha' \, \kappa_D}{\sqrt{D-2}} 
\, \epsilon_g^{\mu_i \nu_i} \, \frac{q_{\mu_i} q_{\nu_i} q_\alpha q_\beta}{k_i \cdot q} \,  \epsilon_g^{\alpha \beta} M_n(g_{\alpha \beta} (k_i+q), \ldots ) \, ,
\nonumber \\[5mm]
M_{n+1}^{\rm ex.}&(\phi(q), B_{\mu_i\nu_i}(k_i), \ldots ) 
 \\[5mm]
\sim&\
\frac{M_{dBB}}{(k_i + q)^2} \, M_n(B_{\alpha \beta}(k_i+q), \ldots ) 
=\frac{\kappa_D}{\sqrt{D-2}} \, \epsilon_B^{\mu_i \nu_i } \frac{ 4 q_{\nu_i} q_{\beta} \eta_{\mu_i \alpha}}{k_i\cdot q}\, \epsilon_B^{\alpha \beta} M_n (B_{\alpha \beta}(k_i+q), \ldots ) \, .
}
These expressions match through order $\Ord(q)$ exactly the singular terms in the dilaton soft theorem, found in the previous section. Specifically, in Eq.~\eqref{softdilatonpoles} (where a factor $\kappa_D/\sqrt{D-2}$ from Eqs.~\eqref{softexpansion} and \eqref{finalsubsubdilaton} was suppressed) one should make the identifications:
\sea{ 
M_n^{\{\mu_i,\nu_i\}} &\equiv \epsilon_g^{\mu_i \nu_i} M_n (g_{\alpha \beta} (k_i), \ldots ) + \epsilon_d^{\mu_i \nu_i} M_n (\phi(k_i), \ldots ) \, , \\
M_n^{[\mu_i,\nu_i]} &\equiv \epsilon_B^{\mu_i \nu_i} M_n (B_{\alpha \beta} (k_i), \ldots ) \, .
}
Notice that the three-point amplitude proportional to $\alpha'$, involving 
one dilaton and two gravitons, does not contribute to the dilaton soft 
theorem, since it is proportional to the fourth power in the soft
 momentum, and thus contributes at the order $q^3$. 

Let us now consider  the
$\alpha'$-correction terms  to the graviton soft theorem 
appearing at subsubleading order in Eq.~\eqref{softgraviton}.
Among them  there is a term with the propagator-pole 
$1/k_i \cdot q$, which should come from a factorizing exchange diagram.
Indeed, expanding the three-point amplitude in terms of $q$, with the 
soft-state now being a graviton, the leading string-correction to the three 
point amplitude reads:
\seq{
M_3^{\mu \nu, \alpha \beta}\Big |_{\alpha'} &\sim - (2 \kappa_D) 
\frac{\alpha'}{2} k_i^\mu k_i^\nu ( (\epsilon_i \cdot q)q^\alpha
 \bar{\epsilon}^\beta + (\bar{\epsilon}_i \cdot q) q^\beta \epsilon_i^\alpha )
 + \Ord(q^3) \\
&=  - (2 \kappa_D) \frac{\alpha'}{2} k_i^\mu k_i^\nu q_\rho q_\sigma 
(\eta^{\sigma \alpha} \epsilon_i^\rho \bar{\epsilon}_i^\beta + 
\eta^{\sigma \beta} \epsilon_i^\alpha \bar{\epsilon}_i^\rho )+ \Ord(q^3) \, ,
\label{M3graviton}
}
where we contracted with the polarization vectors of leg $i$. 
Comparing with Eqs.~\eqref{softgraviton} and \eqref{gravitonstringpole}, we see that the singular string correction in Eq.~\eqref{softgraviton} is exactly reproduced by the exchange diagram given by $\frac{M_3}{(k_i + q)^2}M_n$, with the singularity coming from the pole of the exchanged state.

We will now show that the remaining factorizing non-singular terms in Eq.~\eqref{subsub} all follow 
from on-shell gauge invariance of the amplitude.

\begin{figure}[ht]
\begin{center}
\includegraphics[width=.9\textwidth]{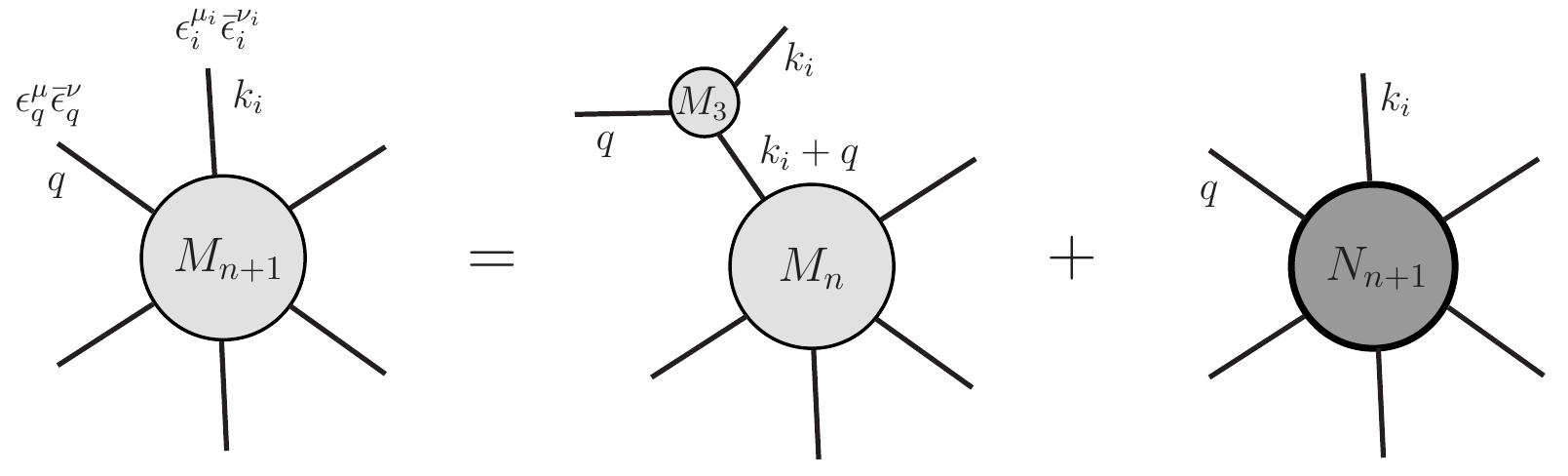}
\caption{Decomposition of the $n+1$-point massless closed string amplitude, 
$M_{n+1}$, into a factorizing part involving the three-point amplitude $M_3$, 
where the soft state with momentum $q$ directly interacts with the $i$th state, 
exchanging a third massless closed state with momentum $k_i + q$ with
 the $n$-point amplitude $M_n$, and the reminding part of the amplitude, 
$N_{n+1}$, which excludes factorization through the former channel.
}
\label{stringfactorization}
\end{center}
\end{figure}

By using the expression for the three-point amplitude in 
Eq.~\eqref{threepoint}, specialized to gravitons in the field theory
 limit $\alpha' \to 0$, it has been shown in Ref.~\cite{BDDN} that by 
decomposing the $n+1$ point amplitude as in Fig.~\ref{stringfactorization}, 
and imposing gauge invariance, i.e. $q_\mu M_{n+1}^{\mu \nu} = 
q_\nu M_{n+1}^{\mu \nu}  = 0$, one exactly recovers the tree-level 
soft graviton theorem to subsubleading order. In the previous sections 
we have explicitly shown to subsubleading order and for a finite value 
of $\alpha'$ (see also Ref.~\cite{DiVecchia:2015oba}) that the same 
gauge invariance also applies to amplitudes where the external states 
can, as well as gravitons, be dilatons and Kalb-Ramond states. 
Based on this, it was shown in Ref.~\cite{DiVecchia:2015jaq} that 
in the limit $\alpha' \to 0$ the dilaton soft theorem to subsubleading 
order can be found also using just gauge invariance of the on-shell amplitude.
 (In Ref.~\cite{DiVecchia:2015jaq} only the cases where the other hard states 
are either dilatons or gravitons were considered, however, the extension to 
also include the antisymmetric Kalb-Ramond states trivially yields the 
expression found in Eq.~\eqref{finalsubsubdilaton}.) Now we extend this 
line of analysis to also involve the string corrections to the three-point 
amplitude.

We first decompose the $n+1$-point amplitude as
 in Fig.~\ref{stringfactorization}, reading:
\ea{
M_{n+1}^{\mu \nu} = \sum_{i=1}^n M_3^{\mu \nu}(q; k_i) \frac{1}{(k_i+q)^2}
M_n(k_i+q) + N^{\mu \nu} (q; k_i) \, ,
}
where dependence on all other $k_j \neq k_i$ is implicit. The indices 
$\mu \nu$ belong to the soft state with momentum $q$, and we 
assume them to be symmetric, i.e. $M_{n+1}^{\mu \nu} = M_{n+1}^{\nu \mu}$. 
We now use the explicit form of the three-point string amplitude, $M_3$, and 
focus here only on the $\alpha'$-terms.
We can read off  these terms, when $\mu \nu$ is symmetric, from 
Eq.~\eqref{M3graviton}, and as we noticed before, they can be 
rewritten as an operator by using Eq.~\eqref{gravitonstringpole}, when 
they are multiplied $M_n$. Thus we have:
\ea{
M_{n+1}^{\mu \nu}\Big |_{\alpha'} =
- \frac{\alpha'}{2} \sum_{i=1}^n \frac{k_i^\mu k_i^\nu }{k_i \cdot q}
\, q_\rho q_\sigma \, \Pi_i^{\{\rho, \sigma\}}\, 
M_n(k_i) + N^{\mu \nu} (q; k_i)\Big |_{\alpha'} \, + \Ord(q^2)\, ,
\label{stringterms}
}
Now, imposing on-shell gauge invariance of the amplitude, we first find from $q_\mu M_{n+1}^{\mu \nu}=0$, and Taylor expanding in $q$:
\sea{
N^{\mu \nu} (q=0, k_i) \Big |_{\alpha'} &= 0 \, ,\\[2mm]
\frac{1}{2} \left (
\frac{\partial N^{\mu \nu}}{\partial q_\rho} + 
\frac{\partial N^{\rho \nu}}{\partial q_\mu} \right ) \Big |_{\alpha',\,  q=0}
&= \alpha' \sum_{i=1}^n k_i^\nu \Pi_i^{\{\rho, \mu\}} M_n(k_i) \, .
}
Inserting this back into the Taylor-expanded form of Eq.~\eqref{stringterms}, and imposing also $q_\nu M_{n+1}^{\mu \nu}=0$, we get:
\ea{
\frac{1}{2} \left (
\frac{\partial N^{\mu \nu}}{\partial q_\rho} - 
\frac{\partial N^{\rho \nu}}{\partial q_\mu} \right )\Big |_{\alpha', \, q=0}
= \frac{\alpha'}{2} \sum_{i=1}^n \left ( 
k_i^\mu \Pi_i^{\{\rho, \nu\}} 
- k_i^\nu \Pi_i^{\{\rho, \mu\}} 
-k_i^\rho \Pi_i^{\{\mu, \nu\}} \right ) M_n(k_i) \, .
}
Thus, in total we find:
\ea{
M_{n+1}^{\mu \nu}\Big |_{\alpha'} =
- \frac{\alpha'}{2} q_\rho \sum_{i=1}^n 
\left (
\frac{k_i^\mu k_i^\nu }{k_i \cdot q}
\, q_\sigma \, \Pi_i^{\{\rho, \sigma\}}
-
  k_i^\mu \Pi_i^{\{\rho, \nu\}}
 - 
  k_i^\nu \Pi_i^{\{\rho, \mu\}}
 + k_i^\rho \Pi_i^{\{\mu, \nu\}}
 \right )
M_n(k_i) + \Ord(q^2) \, ,
}
which demonstrates that, when the soft state is a graviton or a dilaton, on-shell gauge invariance of the string amplitude implies a soft theorem at subsubleading order even when string corrections are taken into account.
Finally we can rewrite this expression as:
\ea{
M_{n+1}^{\mu \nu}\Big |_{\alpha'} =
\frac{\alpha'}{2}  \sum_{i=1}^n 
\left ( 
q_\rho  k_i^\mu \eta^{\nu}_\sigma 
 +
  q_\sigma k_i^\nu \eta^{\mu}_\rho
 - (k_i \cdot q) \eta^{\mu}_{\rho} \eta_\sigma^\nu
 -\frac{k_i^\mu k_i^\nu }{k_i \cdot q}
\, q_\rho q_\sigma
 \right )
 \Pi_i^{\{\rho, \sigma\}}
M_n(k_i) + \Ord(q^2) \, .
}
This is exactly the expression we found in Eq.~\eqref{generalsubsub}
 from explicit calculations. As we noticed in Eq.~\eqref{finalsubsubdilaton}, 
this expression vanishes when contracted with the polarization of the dilaton. 
For the graviton, we have explained that the term singular in $q\to 0$ comes 
from the exchange diagram where the the soft graviton scatters on an external 
massless closed string. When that string is a graviton, respectively, dilaton, the 
exchanged states is the opposite, i.e. a dilaton, respectively, graviton, meaning 
that the $n$-point amplitude in the soft theorem may involve other external 
hard states, than those specified in the $n+1$-point amplitude. 
The non-singular terms on the other hand are, as we have shown, direct 
consequences of on-shell gauge invariance.

For the sake of completeness we conclude this section by comparing to the low-energy
effective action that appears in Ref.~\cite{Metsaev:1987zx}, given by
\seq{
S = \int d^D x \sqrt{-G} &\left\{ \frac{1}{2 \kappa_D^2} R - \frac{1}{2} 
 \partial_{\mu}
\phi  \,\, G^{\mu \nu} \partial_{\nu} \phi   - \frac{1}{24 \kappa_D^2
} {\rm e}^{- 
\frac{4 \kappa_D \phi }{\sqrt{D-2}} } H^2 \right.  
\\
&  \left.
+ \frac{\alpha'}{4} {\rm e}^{- \frac{2 \kappa_D  \phi}{\sqrt{D-2} } } 
\left[ \frac{1}{2 \kappa_D^2} \left( R^2_{\mu \nu \rho \sigma}
- 4 R_{\mu \nu}^2 + R^2 \right) 
+ \cdots
\right]  + \Ord(\alpha'^2) \right\} \, ,
\label{MTcanonical}
}
where, with respect to Eq.~(3.1) of Ref.~\cite{Metsaev:1987zx}, we have chosen a 
different overall 
normalization and have introduced a canonically normalized dilaton. The
first line corresponds to the field theory limit and from it we can reproduce 
the two first couplings in Eq.~(\ref{3couplings}) together with the field theoretical
three-graviton amplitude.
The second line shows the string corrections linear in $\alpha'$, where we have only written down the terms that gives corrections to the three-point amplitude, while the ellipsis $\cdots$ denotes the higher-point operators.
In particular, the last coupling in Eq.~(\ref{3couplings}) is reproduced 
by taking the lowest term of the Gauss-Bonnet part together with a dilaton coming from
the exponential in front of the Gauss-Bonnet term, while the cubic term 
in the metric coming from only the Gauss-Bonnet part reproduces the first
string correction to the three-graviton amplitude in Eq.~(\ref{threepoint}), 
as already noticed in Ref.~\cite{Zwiebach:1985uq}.

\section{Discussion and Conclusion}
\label{conclusion}

In this paper we completed the computation of the amplitude involving $n+1$ massless closed string states in the bosonic string to the subsubleading order in the soft momentum expansion of one of the external states. When the soft state is either a graviton or a dilaton we showed that the result can be expressed as a soft theorem to subsubleading order. The graviton soft theorem has string corrections at the subsubleading order in the bosonic string, while the string corrections for the dilaton all vanish. The leading and subleading terms
were already obtained in Ref.~\cite{DiVecchia:2015oba}.

The calculation involves an extension of the technique developed in 
Ref.~\cite{DiVecchia:2015oba}, which is rather involved, but the final result has a very simple explanation:
The basic ingredients to derive the soft-theorems are gauge invariance and the three-point amplitude with massless closed string states that, in the bosonic string, contains also string corrections.  The procedure is an extension of the one followed in 
Ref.~\cite{BDDN}, where the total amplitude is written as a sum of two types of terms.
One corresponds to diagrams where the soft leg is attached to each of the
other external legs through the three-point amplitude of the soft state 
and two other massless states. These terms trivially factorize and in general have a pole when the momentum of the soft state goes to zero. The other type of terms do not factorize trivially and are finite in the soft limit. The two types of contributions, however, are not separately gauge invariant.
Thus, by imposing on-shell gauge invariance of the full amplitude, one is able to
factorize the soft behavior of the total amplitude up to subsubleading order in 
the soft momentum. With respect to the procedure of Ref.~\cite{BDDN} for the
case of a soft graviton, there are, however, two important differences.  
Gauge invariance
imposed on the amplitude $M_{\mu \nu}$ does not only give the soft behavior
for the graviton, when saturated with the graviton polarization 
$\epsilon^{\mu \nu}_g$, but also that for the dilaton when saturated with the
polarization of the dilaton $\epsilon^{\mu \nu}_d$. The second difference with
respect to the field theoretical behavior of an amplitude with only gravitons,
is that, in general, the state exchanged  in the propagator of the pole term
is not necessarily the same that appears in the corresponding external leg.
This happens in the string correction term in the amplitude with only gravitons 
and in the case of a soft dilaton. 

We thus arrive to the conclusion that the soft theorems for 
the graviton and the dilaton are both consequence of gauge invariance and 
that the string corrections appearing at subsubleading order for the graviton are direct consequences of the three-point amplitude with massless closed strings 
having string corrections in the bosonic string. Since the three-point amplitude in superstring
has no string corrections, we expect no string corrections in the soft behavior
of a graviton in superstring. 

A curious outcome of our results is that the soft theorem of a dilaton contains, at subleading
order, the generator of scale transformation and, at subsubleading order, the
generator of special conformal transformations as is also the case in the 
soft theorem of a Nambu-Goldstone boson of spontaneously broken 
conformal symmetry~\cite{DiVecchia:2015jaq}. Some discussions of this was given in Ref.~\cite{DiVecchia:2015jaq}, but it would be interesting to have a more physical 
understanding of why this happens.   Finally, it would also be interesting to 
extend our considerations to the soft behavior of the Kalb-Ramond 
antisymmetric tensor, and furthermore how the results go over to the case of superstrings.

\vspace{-5mm}
\subsection*{Acknowledgments} \vspace{-3mm}
We thank Massimo Bianchi, Marco Bill{\`{o}},  Marialuisa Frau, Andrea Guerrieri, 
Alberto Lerda, Carlo Maccaferri,  Josh Nohle, Igor Pesando and  Congkao Wen for many useful discussions. We owe special thanks to Josh Nohle for many relevant comments 
on our results.

\appendix
\section{Results of integrals}
\label{Results}
All integrals necessary for the results of this paper can  be reduced to the general form:
\ea{
I_{i_1 i_2 \ldots}^{j_1 j_2 \ldots} =
\frac{1}{2\pi}
\int d^2 z \frac{\prod_{l = 1}^n |z-z_l|^{\alpha' k_l q}}{
(z-z_{i_1})(z-z_{i_2}) \cdots (\bar{z}-\bar{z}_{j_1}) (\bar{z}-\bar{z}_{j_2}) \cdots } \ .
\label{GeneralIntegral}
}
We are using the convention $d^2 z = 2 d {\rm Re}(z) d {\rm Im} (z)$. The quantity $I^{i_1 i_2 \ldots}_{j_1 j_2 \ldots}$ is symmetric in its lower,  respectively upper indices and complex conjugation exchanges the lower and upper indices. 
These integrals can generically be evaluated after an expansion in $q$ by a substitution of the form $z \to z_{i} + \rho e^{i\theta}$. 
It can be useful to substitute further $e^{i\theta} \to \omega$, such that  $\int_{0}^{2\pi}  d \theta \cdots \to \oint d \omega \cdots$,  enabling use of Cauchy's integral formula over the unit circle. 
To evaluate the integrals one must consider, case by case, when indices of $I^{i_1 i_2 \ldots}_{j_1 j_2 \ldots}$ are equal or not. Thus, in the following different labels on indices indicates the case when they are not equal to each other.

We have explicitly computed all integrals to the relevant order in $q$ involved in this paper using the mentioned procedure. Through this procedure, in Ref.~\cite{DiVecchia:2015oba} the integrals $I^i_i$ and $I^i_j$ were already computed to the order $q^1$, with the results:
\ea{
I_i^i = & 
 \frac{2}{\alpha'k_i q}  
\left( 1 +\alpha'  \sum_{j \neq i} (k_j q) \log |z_i - z_j|  
 + \frac{(\alpha')^2}{2} 
\sum_{j \neq i} \sum_{k \neq i} (k_j q) (k_k q)  \log|z_i -z_j| \log |z_i - z_k| \right) 
 \nonumber \\
&+(\alpha')^2 \sum_{j \neq i} (k_j q) \log^2 |z_i - z_j| + \log \Lambda^2 + \Ord(q^2)\label{A}
 \\[5mm]
I_i^j = &
\sum_{m\neq i,j}\frac{\aqk{m}}{2}\left ({\rm Li}_2\left( \frac{\bar{z}_i-\bar{z}_m}{\bar{z}_i-\bar{z}_j}\right)-{\rm Li}_2\left(\frac{z_i-z_m}{z_i-z_j }\right)
-2\log\frac{\bar{z}_m-\bar{z}_j}{\bar{z}_i-\bar{z}_j}\log\frac{ |z_i-z_j|}{|z_i-z_m|}
\right )
\nonumber\\
&- \log|z_i-z_j|^2+\log\Lambda^2  + \Ord(q^2)\label{IIj}
}
where $\Lambda$ is a cutoff that vanishes, when inserted in the appropriate expression.

\noindent
It is useful to express the symmetrized and antisymmetrized form of latter expressions:
\ea{
\frac{1}{2}\left(I_{i}^j+I_{j}^i\right)= & \log\Lambda^2 - \log|z_i-z_j|^2+\sum_{m\neq i,j}\aqk{m}
\Bigg (\log|{z}_m-{z}_j|\log|z_i-z_m|\nonumber\\
&- \log|{z}_m-{z}_j|\log|z_i-z_j|-\log|{z}_i-{z}_j|\log|z_i-z_m| 
\Bigg ) + \Ord(q^2)
\\[5mm]
\frac{1}{2}\left(I_{i}^j-I_{j}^i\right)=&\frac{1}{2}\sum_{m\neq i}\alpha' q\cdot k_m{\rm Li}_2\left( \frac{\bar{z}_i-\bar{z}_m}{\bar{z}_i-\bar{z}_j}\right)-\frac{1}{2} \sum_{m\neq i}\alpha' q\cdot k_m{\rm Li}_2\left(\frac{z_i-z_m}{z_i-z_j }\right)\nonumber\\
&+\frac{1}{2}\sum_{m\neq i,j}\alpha' q\cdot k_m \log\frac{|z_i-z_j|}{|z_i-z_m|}\log\left[\frac{z_m-z_j}{\bar{z}_m-\bar{z}_j}
\frac{\bar{z}_i-\bar{z}_j}{z_i-z_j}\right] + \Ord (q^2)
}

It is a long and tedious exercise to evaluate all other integrals by the above mentioned procedure.
In this appendix we therefore instead present an alternative shorter derivation leading to the same results, by making use of the above two expressions.

By using partial derivatives with respect to the $z_i$'s it is possible to directly get the higher-index integrals from the two basic integrals above. The following four integrals for instance are readily derived in this way, and they will be used, together with the previous two integrals, as master integrals as will be explained below:
\ea{
 {I_{ii}^i} =&\, \frac{1}{1-\frac{\alpha'}{2}k_i q}\partial_{z_i}I_{i}^i=
\sum_{j\neq i} 
\Bigg \{
\frac{1}{z_i-z_j}
\Bigg[
\frac{ q k_j}{qk_i}
+ \alpha' q \cdot k_j \LN{i}{j}
+\frac{\alpha' (q k_j) }{2} 
\Bigg]
\nonumber \\
&+\,
\frac{1}{2}\sum_{l\neq i} \frac{(\alpha' q k_j)(\alpha' q k_l)}{\alpha' q k_i } 
\left [ 
\frac{\log |z_i - z_j|}{z_i - z_l} + \frac{\log |z_i - z_l|}{z_i - z_j}\right ]
\Bigg \}
+ \Ord(q^2) \label{Ii-ii}\\[5mm]
 {I_{ii}^j} =&\, \frac{1}{1-\frac{\alpha'}{2} 
(k_i q)} \partial_{z_i}I_{i}^{j}=
- \frac{1}{z_i - z_j } \left (
1 + \frac{\alpha' q k_i  }{2} 
+ \alpha' (q k_i+q k_j) \LN{i}{j}
\right )
 \nonumber \\
 &
+  \sum_{l \neq i,j} \alpha' q k_l
\Bigg\{ 
\frac{\log \left | z_j-z_l \right |}{z_i - z_l} 
-\frac{\log \left | z_i-z_j \right |}{z_i - z_l} 
-
\frac{\log |z_j-z_l |}{z_i - z_j}
\Bigg \} + \Ord(q^2)\label{Ij-ii} 
}
\ea{
I_{ii}^{ii} =&\, \frac{1}{1-\frac{\aqk{i}}{2}} \partial_{\bar{z}_i} I_{ii}^i 
\nonumber \\
=&\, \frac{1}{2} 
\sum_{j \neq i} 
\frac{\aqk{j}}{|\zz{i}{j}|^2} 
\Bigg (1
+ \frac{\qk{j}}{\qk{i}}
+\frac{1}{2}\sum_{l\neq i,j}
 \frac{\qk{l}}{\qk{i}}
\left [
\frac{\zbzb{i}{j}}{\zbzb{i}{l}}+
\frac{\zz{i}{j}}{\zz{i}{l}} \right ]
\Bigg )
+ \Ord(q^2) \label{Iii-ii}\\[5mm]
I_{ii}^{jj} =&\, \frac{1}{1-\frac{\aqk{j}}{2}} \partial_{\bar{z}_j} I_{ii}^j = 
-\frac{1}{2|\zz{i}{j}|^{2}} 
\sum_{l \neq i,j} \aqk{l} 
\frac{(\zbzb{i}{l})(\zz{j}{l})}{(\zz{i}{l})(\zbzb{j}{l})} 
+ \Ord(q^2)
\label{Iii-jj}
}
All the other integrals relevant in the soft limit  are related to the six master integrals by simple algebraic relations which follows from a useful   rewriting of the holomorphic and anti-holomorphic poles  of Eq.~(\ref{GeneralIntegral}). $I^{j_1j_2\dots}_{i_1 i_2\dots}$ has  poles both in $z=z_{i_1}$, $z=z_{i_2}$ and in  $\bar{z}=\bar{z}_{j_1}$, $\bar{z}=\bar{z}_{j_2}$. It can be   written as linear combination of the quantities $I^{j_1\dots}_{i_1\dots}$, $I^{j_2\dots}_{i_1\dots}$, $I^{j_1\dots}_{i_2\dots}$ $I^{j_2\dots}_{i_2\dots}$ having only two of the above mentioned poles. This follows from the  identity:
\begin{eqnarray}
\frac{1}{(z-z_{i_1})(z-z_{i_2})} = \frac{1}{(z_{i_1}-z_{i_2})}\left[\frac{1}{(z-z_{i_1})}-
\frac{1}{(z-z_{i_2})}\right]
\end{eqnarray}
with a similar relation for the $\bar{z}$ variable. This determines:
\begin{eqnarray}
 I^{j_1j_2\dots}_{i_1 i_2\dots}=\frac{ I^{j_1j_2\dots}_{i_1 \dots}-I^{j_1j_2\dots}_{ i_2\dots}}{(z_{i_1}-z_{i_2})}= \frac{ I^{j_1\dots}_{i_1 \dots}-I^{j_2\dots}_{i_1 \dots}-I^{j_1\dots}_{ i_2\dots}+I^{j_2\dots}_{ i_2\dots}}{(z_{i_1}-z_{i_2}) (\bar{z}_{j_1}-\bar{z}_{j_2})}\label{idm}
\end{eqnarray}
By applying  Eq.~(\ref{idm}) iteratively, all the necessary integrals can be computed.
Following such an analysis, we present in the subsequent sections the results of all integrals involved in this work to the necessary order in $q$.
It will be useful to notice from the above decomposition that an integral $I^{j_1j_2\dots}_{i_1 i_2\dots}$ can yield a term of order $q^{-1}$ only if one of its lower indices is equal to one of the upper ones, since its decomposition may only then contain $I_{i}^i$. This property will be used throughout our derivations.

\subsection*{Integrals of the form $I_{im}^{jn}$}
The integrals $I_{ii}^{ii}$ and $I_{ii}^{jj}$ were already given above in Eqs.~\eqref{Iii-ii}-\eqref{Iii-jj}.

Next, according to  Eq.~(\ref{idm}) we have:
\ea{
I_{ii}^{ij}= \frac{I^i_{ii}-I^j_{ii}}{(\bar{z}_i-\bar{z}_j)}  = \frac{1}{|z_i-z_j|^2}  \left (1+  \frac{\qk{j}}{\qk{i}} + \sum_{l\neq i,j} \frac{\qk{l}}{\qk{i}} \frac{\zz{i}{j}}{\zz{i}{l}} \right )\label{Iii-ij}
+\Ord(q)
}
The quantities $I^i_{ii}$ and $I^j_{ii}$ have been computed up to the order  $q$  and therefore we could give $I_{ii}^{ij}$ up to this order in the soft expansion. However, this integral appears in $S_1$ and $S_3$ multiplied by a factor $q$, therefore, only the terms of order $q^0$ are relevant.  
This kind of restriction will be implicit in all the following results.

Then,
\ea{
  {I_{ii}^{jm}}=&\frac{I^j_{ii}-I^m_{ii}}{\bar{z}_i-\bar{z}_m}=\frac{1}{\zbzb{j}{m}}\left(\frac{1}{z_i-z_m}-\frac{1}{z_i-z_j}
 \right)+\Ord(q)
 }
\ea{
I_{im}^{im} = \frac{  I^i_i-I^m_i-I^i_m+I^m_m}{|z_i-z_m|^2}=
\frac{2}{|\zz{i}{m}|^2} \left (\frac{1}{\aqk{i}} + \frac{1}{\aqk{m}} \right ) + \Ord(q^0)
}
\ea{
{I_{im}^{in}}= \frac{ I^i_i-I^n_i-I^i_m+I^n_m}{(z_i-z_m)(\bar{z}_i-\bar{z}_n)} 
= \frac{2 }{\aqk{i}}\frac{1}{(\zz{i}{m})(\zbzb{i}{n})} + O (q^0)
}
\ea{
I_{im}^{jn} &=\frac{  I^j_i-I^n_i-I^j_m+I^n_m}{(z_i-z_m)(\bar{z}_j-\bar{z}_n)}= 0 + \Ord(q^0)\label{Iim-jn}
}
As we have already noticed, the expansion of the integrals may start from order $q^{-1}$ only if one of the upper indices is equal  to one of the lower ones. Thus the result of Eq.~\eqref{Iim-jn} could have been readily inferred from this property, without making the intermediate decomposition. This example serves as a demonstration of this property, which we will directly use for the rest of this appendix.

\subsection*{Integrals of the form $I_{imn}^{j}$}
Using Eq.~(\ref{idm}) iteratively, we get
\ea{
I_{iim}^i =&\,\frac{I^i_{ii}}{z_i-z_m}-\frac{ I^i_i-I^i_m}{(z_i-z_m)^2}=-\frac{2}{(z_i-z_m)^2} 
\nonumber\\ 
 &\,\times
 \left [
  \frac{1}{\aqk{i}}
  + \LN{i}{m}
  +
 \sum_{l\neq i}\frac{\qk{l}}{\qk{i}}
\left (
\LN{i}{l}
- \frac{1}{2} \frac{\zz{i}{m}}{\zz{i}{l}}
\right )
\right ]  + \Ord(q) \label{Iiim-i}\\[5mm]
I_{iij}^j =&\,\frac {I^j_{ii}}{z_i-z_j}-\frac{  I^j_i-I^j_j}{(z_i-z_j)^2}\nonumber\\
=&\,  
\frac{2}{(z_i-z_j)^2}
\left [
\frac{1}{\alpha' q k_j} - \frac{1}{2} + \LN{i}{j} +
 \sum_{l\neq j} \frac{\qk{l}}{\qk{j}} \LN{l}{j}
 \right ]+ \Ord(q) \label{Iiij-j} \\[5mm]
 I_{imn}^{i} =&\,\frac{1}{z_m-z_n}\left( \frac{I^i_i-I^i_m}{z_i-z_m}-\frac{I^i_i-I^i_n}{z_i-z_n}\right)
=\frac{2 }{\aqk{i}} \frac{1}{(\zz{i}{m})(\zz{i}{n})} + \Ord(q^0) 
 }
 \ea{
I_{iim}^j =&\,\frac{I^j_{ii}}{z_i-z_m}-\frac{ I^j_i-I^j_m}{(z_i-z_m)^2}\nonumber\\
=&\,- \frac{1}{ (\zz{i}{m})(\zz{i}{j})} +  \frac{2 }{(\zz{i}{m})^2} \left ( \log |\zz{i}{j}| - \log |\zz{j}{m}| \right)
+ \Ord(q) \\[5mm]
I_{imn}^{j} =&\, 0 + \Ord(q^0)
}
The result of the last integral follows immediately due to non of the upper and lower indices being pairwise equal.

\subsection*{Integrals of the form $I_{imn}^{jl}$}

Using Eq.~(\ref{idm}) iteratively, and the identities $I_m^{ii}=(I^m_{ii})^*$ and $I_i^{ii}=(I^i_{ii})^*$, where $*$ denotes  the complex conjugate, we get
\ea{
I_{iim}^{ii} = &\, \frac{I^{ii}_{ii}}{z_i-z_m} -\frac{(I^i_{ii})^* -(I^m_{ii})^*}{(z_i-z_m)^2}
\nonumber \\
= &\, 
-\frac{1}{|\zz{i}{m}|^{2} (\zz{i}{m}) }
\left(1+ \frac{\qk{m}}{\qk{i}}+  \sum_{l\neq i,m} \frac{\qk{l}}{\qk{i}}
\frac{\zbzb{i}{m}}{\zbzb{i}{l}} \right ) + \Ord(q)
\label{Iiimii}
} 
\ea{
I_{iij}^{jj} = &\,  \frac{I^{jj}_{ii}}{z_i-z_j} -\frac{I^{jj}_i-I^{jj}_j}{(z_i-z_j)^2}
\nonumber \\
= &\, 
-\frac{1}{|\zz{i}{j}|^{2} (\zz{i}{j})} \left (1+ \frac{\qk{i}}{\qk{j}} + \sum_{l \neq i,j} \frac{\qk{l}}{\qk{j}}
\frac{\zbzb{j}{i}}{\zbzb{j}{l}}\right ) + \Ord(q)
}
\ea{
{I_{iim}^{jj}} =-\frac{I^{jj}_{ii}}{z_i-z_m} -\frac{I^{jj}_i-I^{jj}_m}{(z_i-z_m)^2}
&=
-  \frac{(\zbzb{i}{m})}{(\zbzb{i}{j}) (\zz{i}{m})^2(\zbzb{j}{m}) } + \Ord(q)
}
\ea{
I_{iij}^{ij} =&\, \frac{I^i_{ii}-I^j_{ii}}{|z_i-z_j|^2} -\frac{ I^i_i-I^i_j-I^j_i+I^j_j}{(z_i-z_j)|z_i-z_j|^2}\nonumber\\
=&\,
-\frac{2}{(\zz{i}{j})|\zz{i}{j}|^2} \left (\frac{1}{\aqk{i}} + \frac{1}{\aqk{j}} \right )
+ \Ord(q^0)
}
\ea{
I_{iik}^{im} = &\,\frac{I^i_{ii}-I^m_{ii}}{(z_i-z_k)(\bar{z}_i-\bar{z}_m)} -\frac{I^i_i-I^i_k+I^m_i-I^m_k}{(z_i-z_k)^2(\bar{z}_i-\bar{z}_m)}\nonumber\\
=&\,
- \frac{2}{(\zz{i}{k})^2 (\zbzb{i}{m}) \aqk{i}} + \Ord(q^0)
}
\ea{
I_{iij}^{jl}=&\,\frac{I^j_{ii}-I^l_{ii}}{(z_i-z_j)(\bar{z}_j-\bar{z}_l)}-\frac{I^j_i-I^j_j+I^l_i-I^l_j}{(z_i-z_j)^2(\bar{z}_j-\bar{z}_l)}= \frac{2}{\aqk{j}}\frac{1}{(\zz{i}{j})^2 (\zbzb{j}{l})} + \Ord(q^0)
}
Finally, the integrals $I_{iim}^{jn}$, $I_{imn}^{ii}$, $I_{imn}^{jj}$ have either all the upper indices different from the lower ones or more than one lower index equal to an upper one. Thus according to the general property, noticed specifically after Eq.~(\ref{Iim-jn}),  we have: 
\ea{
I_{iim}^{jn} =I_{imn}^{ii} = I_{imn}^{jj} =0 + \Ord(q^0)
\label{Iiimjn}
}

\subsection*{Integrals of the form  $I_{iimn}^{j}$}

Using Eq.~(\ref{idm}) iteratively,
\ea{
I_{iikm}^i =&\, \frac{1}{(z_k-z_m)(z_i-z_k)}\left (I^i_{ii} -\frac{I^i_i-I^i_k}{(z_i-z_k)}\right ) -\frac{1}{(z_i-z_m)(z_k-z_m)}\left (1-\frac{I^i_i-I^i_m}{(z_i-z_m)} \right )\nonumber\\
=&\, \frac{2}{\aqk{i}}\frac{1}{(\zz{m}{k})} \left [
\frac{1}{(\zz{i}{k})^2} 
- \frac{1}{(\zz{i}{m})^2} 
\right ] + \Ord(q^0)
}
\ea{
I_{iijl}^j =&\, \frac{I^j_{ii}}{(z_i-z_l)(z_i-z_j)} -\frac{I^j_i-I^j_j}{(z_j-z_l)(z_i-z_j)^2}-\frac{I^j_{ii}}{(z_j-z_l)(z_i-z_l)}+\frac{I^j_i-I^j_l}{(z_j-z_l)(z_i-z_l)^2}\nonumber\\
=&\, \frac{2}{\aqk{j}}\frac{1}{(\zz{j}{i})^2(\zz{j}{l})} + \Ord (q^0)
}
\ea{
I_{iimn}^j = 0 + \Ord(q^0)
}
The last result follows since none of the lower indices are equal to the upper one.

\subsection*{Integrals of the form $I_{imn}^{jjn}$}
Using Eq.~(\ref{idm}) iteratively,
\ea{
I_{iik}^{iik} =&\frac{I^{ii}_{ii}}{|z_i-z_k|^2}   -\frac{I^i_{ii}-I^k_{ii}}{|z_i-z_k|^2(\bar{z}_i-\bar{z}_k)}-\frac{I^{ii}_i-I^{ii}_k}{(z_i-z_k)|z_i-z_k|^2}+\frac{I^i_i-I^k_i-I^i_k+I^k_k}{|z_i-z_k|^4}\nonumber\\
=&
\frac{2}{|\zz{i}{k}|^4} \left (\frac{1}{\aqk{i}}+\frac{1}{\aqk{k}}\right )
 + \Ord (q^0)
}
\ea{
I_{iik}^{iim}
=&\frac{I^{ii}_{ii}}{(z_i-z_k)(\bar{z}_i-\bar{z}_m)}-\frac{I^i_{ii}-I^m_{ii}}{(z_i-z_k)(\bar{z}_i-\bar{z}_m)^2}-\frac{I^{ii}_i-I^{ii}_k}{(z_i-z_k)^2(\bar{z}_i-\bar{z}_m)}+\frac{I^i_i-I^i_k-I^m_i+I^m_k}{(z_i-z_k)^2(\bar{z}_i-\bar{z}_m)^2}\nonumber\\
=&
\frac{2}{\aqk{i} (\zz{i}{k})^2(\zbzb{i}{m})^2} + \Ord(q^0)
}
\ea{
I_{iij}^{jji} &=\frac{I^{jj}_{ii} }{|z_i-z_j|^2} -\frac{I^j_{ii}-I^i_{ii}}{|z_i-z_j|^4(\bar{z}_i-\bar{z}_j)}-\frac{I^{jj}_i-I^{jj}_j}{|z_i-z_j|^2(z_i-z_j)}+\frac{I^j_i-I^i_i-I^j_j+I^i_j}{|z_i-z_j|^4}\nonumber\\ 
=&- \frac{2}{|\zz{i}{j}|^4} \left (\frac{1}{\aqk{i}}+\frac{1}{\aqk{j}} \right ) + \Ord(q^0)
}
\ea{
I_{iik}^{jjk}=& \frac{I^{jj}_{ii}}{(z_i-z_k)(\bar{z}_j-\bar{z}_k)}-\frac{I^j_{ii}-I^k_{ii}}{(z_i-z_k)(\bar{z}_j-\bar{z}_k)^2}-\frac{I^{jj}_i-I^{jj}_k}{(z_i-z_k)^2(\bar{z}_j-\bar{z}_k)}+\frac{I^j_i-I^j_k-I^k_i+I^k_k}{(z_i-z_k)^2(\bar{z}_j-\bar{z}_k)^2}\nonumber\\
=&
\frac{2}{(\zbzb{k}{j})^2 (\zz{k}{i})^2} \frac{1}{\aqk{k}} + \Ord(q^0)
}
\ea{
I_{iik}^{jji} &=\frac{I^{jj}_{ii} }{(z_i-z_k)(\bar{z}_j-\bar{z}_i)} -\frac{I^j_{ii}-I^i_{ii}}{(z_i-z_k)(\bar{z}_j-\bar{z}_i)^2}-\frac{I^{jj}_i-I^{jj}_k}{(z_i-z_k)^2(\bar{z}_j-\bar{z}_i)}+\frac{I^j_i-I^j_k-I^i_i+I^i_k}{(z_i-z_k)^2(\bar{z}_j-\bar{z}_i)^2}\nonumber\\
=&
- \frac{2}{(\zz{i}{k})^2(\zbzb{i}{j})^2 \aqk{i}} + \Ord(q^0)
} 
\ea{
I_{iim}^{jjn} = 0 + \Ord(q^0)
}
The last result follows since none of the lower indices are equal to the upper ones.

\subsection*{Integrals of the form  $I_{iimn}^{jj}$}

For these integrals either two upper indices are equal to some lower indices, or 
none are equal. Thus neither of the cases contains $I_i^i$ in its decomposition and thus in any case,
\ea{
I_{iimn}^{ii} = I_{iijm}^{jj} =I_{iimn}^{jj}  =0 + \Ord(q^0)\label{A88}
}

\section{Matching with the subsubleading soft operator}
\label{AppB}

In this appendix we explicitly show how our main result in 
Eq.~\eqref{generalsubsub} reproduces our explicit calculation 
of the amplitude of $n+1$ massless closed states in the  bosonic 
string, presented in Sect.~\ref{stringdilaton} and App.~\ref{Results}, when 
the soft state is a graviton or a dilaton.

In Eq. (\ref{generalsubsub}), the subsubleading soft  operator is the sum of 
four parts which act on the $n$-point amplitude. The result  of this action is 
given  
below and compared with explicit results in Eqs.~(\ref{Sl1})-(\ref{Sl3}).  

The action of the first term in Eq.~(\ref{generalsubsub}) on the $n$-point 
amplitude is:
\ea{
-\sum_{i=1}^n\frac{ q_\rho J_i^{\mu\rho}q_\sigma J_i^{\nu\sigma}}{2k_iq}  M_n=&M_n*\left[ 
A^{\mu\nu}_{(\theta^0)}+A^{\mu\nu}_{(\theta)}+A^{\mu\nu}_{(\theta^2)}+
A^{\mu\nu}_{(\theta\bar{\theta})}+ A^{\mu\nu}_{(\theta^3)}
+A^{\mu\nu}_{(\theta^2\bar{\theta} )}+ A^{\mu\nu}_{(\theta^2\bar{\theta}^2)}\right]
+c.c.
\label{adef}
}
We are denoting with $A^{\mu\nu}_{(\theta^\alpha \bar{\theta}^\beta)}$, $\alpha,
\beta=0,1, 2$, terms with different powers  in the Grassmann variables  $(\theta_i,\, 
\bar{\theta}_j)$. 

The action of the second operator in Eq.~(\ref{generalsubsub}) gives:
\ea{
&-\frac{1}{2} \sum_{i=1}^n\left( \frac{k_i^\nu q^\mu}{k_iq}q^\sigma +q^\nu \eta^{\mu\sigma} -\eta^{\nu\mu} q^\sigma\right)\frac{\partial}{\partial k_i^\sigma}M_n=-M_n*\Bigg[\frac{1}{2}\sqrt{\frac{\alpha'}{2} } \sum_{i\neq j} \left( \frac{k_i^\nu q^\mu (k_jq)}{k_iq}\frac{ (\theta_j\epsilon_jq)}{z_j-z_i}\right.\nonumber\\
&\left.+\frac{q^\nu \theta_j\epsilon_j^\mu}{z_j-z_i}-\frac{\eta^{\mu\nu} (\theta_j\epsilon_jq)}{z_j-z_i}+ c.c.\right)-\frac{\alpha'}{2} \sum_{i\neq j} \left(
k_i^\mu q^\nu\frac{k_jq}{k_iq}+q^\nu k_j^\mu -\eta^{\mu\nu} qk_j\right)\log|z_i-z_j|\Bigg].
\label{secondoperator}
}
The action of the operators in the second and third line of Eq.~\eqref{generalsubsub} will be considered in the end.

We restrict our analysis to the case in which the external soft particle 
is a graviton  or a dilaton, meaning that the expressions above are projected on the symmetric combination of $\mu \nu$.
All $A^{\mu\nu}_{(\theta^\alpha \bar{\theta}^\beta)}$ depend on the polarizations and momenta of the hard particles. Their explicit expressions are rather lengthy, but we may express them in more compact terms by using the explicit expressions for the quantities $I_{i_1 i_2 \ldots}^{j_1 j_2 \ldots}$ given in App.~\ref{Results}, which also allows for an easier identification with the expression in Eqs.~(\ref{Sl1})-(\ref{Sl3}). The first quantity can be expressed as:
\ea{
A^{\mu\nu}_{(\theta^0)}= \frac{\alpha'}{4} \left[ 
\sum_{i\neq j}\big(k_i^\nu q^\mu\frac{(k_jq)}{k_iq}-\eta^{\mu\nu} (k_jq) +
k_j^\mu q^\nu\big)
\log|z_i-z_j|+\sum_{i,j=1}^n
k_i^{(\mu } k_j^{\nu)} 
{I^{(1)}}^j_i 
\right] \label{theta0}
}
with $k_i^{\{\mu} k_j^{\nu\}}= \frac{1}{2}(k_i^{\mu} k_j^{\nu}+k_i^\nu k_j^{\mu})$. 
This expression is real and thus the $c.c.$ terms simply give a factor of two.
Because of this factor $2$, the term  with $\log |z_i - z_j|$ in 
Eq.~\eqref{theta0} cancel the analogous term in 
Eq.~\eqref{secondoperator}.
The remaining term in Eq.~\eqref{theta0} involving $I_i^j$ coincides with 
the symmetric part of  $S_1^{(0)}$ in Eq.~\eqref{S1su0giu}.

The next term reads: 
\begin{eqnarray}
&&A^{\mu\nu}_{(\theta)}= \sqrt{\frac{\alpha'}{2}} \Bigg\{\frac{1}{2} \sum_{i\neq j} \left[ 
\left( \eta^{\mu\nu}-\frac{k_i^\nu q^\mu}{k_iq}\right) \frac{(\theta_j\epsilon_jq)}{z_i-z_j} -
\frac{\theta_j\epsilon_j^\mu q^\nu}{z_i-z_j}\right.\nonumber\\
&&\left. +\frac{k_jq}{k_iq} \frac{\theta_i\epsilon_i^\nu q^\mu}{z_i-z_j}+\left(\frac{k_j^\mu 
q^\nu}{qk_i}-  \eta^{\mu\nu}\frac{k_jq}{k_iq}\right)\frac{\theta_i\epsilon_iq}{z_i-z_j}
\right]+\frac{\alpha'}{2} \sum_{i\neq j}\left[ \frac{(k_iq)
\theta_i\epsilon_i^{\{\mu}k_j^{\nu\}} }{z_i-z_j}-\frac{ (k_jq)
\theta_i\epsilon_i^{\{\mu}k_i^{\nu\}}}{z_i-z_j}\right]\nonumber\\
&&-\frac{\alpha'}{2} \sum_{i\neq j} k_i^{\{\mu} k_i^{\nu\}} (\theta_j\epsilon_jq){I^{(0)}}^i_{ij} -\frac{\alpha'}{2}
\sum_{i\neq j}k_i^{\{\mu} k_j^{\nu\}} {I^{(0)}}^j_{ij}(\theta_j\epsilon_jq)-\frac{\alpha'}{2} 
\sum_{i\neq j\neq l} k_i^{\{\mu} k_i^{\nu\}}(\theta_l\epsilon_lq) {I^{(0)}}^j_{il}\nonumber\\
&&+\sum_{i,j=1}^n\theta_i\epsilon_i^{\{\mu}k_i^{\nu\}} {I^{(1)}}^j_{ii}\Bigg\}
\label{theta1}
\end{eqnarray}
The first three terms and their complex conjugate cancel the first three 
terms in Eq.~\eqref{secondoperator} and their complex conjugate, 
coming from the second soft operator.
The third line  of this expression makes up part of $S_1^{(1)}$ when the soft state is taken to be symmetrically polarized, i.e $\epsilon_{q\mu} \bar{\epsilon}_{q\nu} \to {\epsilon_{q \mu\nu}^S}$.
The following terms of $S_1^{(1)}$ then remain to be matched:
\begin{eqnarray}
-\left(\frac{\alpha'}{2}\right)^{\frac{3}{2}} \sum_{i\neq j} (\epsilon_qk_i)
(\bar{\epsilon}_qk_i)(\theta_i\epsilon_iq){I^{(0)}}^i_{ii}
-\left(\frac{\alpha'}{2}\right)^{\frac{3}{2}} \sum_{i\neq j} 
(\epsilon_qk_i)(\bar{\epsilon}_qk_j)(\theta_i\epsilon_iq){I^{(0)}}^j_{ii}
\label{rm1}
\end{eqnarray}
The last term in Eq.~(\ref{theta1})  is equal to $S_3^{(0)}$ when the soft state again is a graviton or a dilaton.
Finally, the second line does not match any term in $S_1$, $S_2$ and $S_3$, but they will be cancelled by the additional soft operators, discussed in the end, which at the same time will also produce the missing terms in Eq.~\eqref{rm1}.

Next, we consider
\begin{eqnarray}
&&A^{\mu\nu}_{(\theta^2)}= 
\frac{1}{2} \sum_{i\neq j}\left[\frac{\theta_i\epsilon_i^\nu q^\mu(\theta_j\epsilon_jq) -(\eta^{\mu\nu} (\theta_j\epsilon_jq) -q^\nu\theta_j\epsilon_j^\mu)(\theta_i\epsilon_iq)}{qk_i(z_i-z_j)^2}\right]
\nonumber \\
&&+\left( \frac{\alpha'}{2} \right)^2\sum_{i\neq j}\frac{k_i^{\{\mu}k_j^{\nu\}} (\theta_i\epsilon_iq)(\theta_j\epsilon_jq)}{(z_i-z_j)^2} {I^{(-1)}}^j_j +\frac{1}{2}\left( \frac{\alpha'}{2} \right)^2\sum_{i\neq j} k_i^\mu k_i^\nu  
\sum_{l\neq i,j} \frac{(q\theta_j\epsilon_j)(q\theta_l\epsilon_l){I^{(-1)}}^i_i}{(z_i-z_j)(z_i-z_l)}
\nonumber\\
&&
-\left( \frac{\alpha'}{2} \right)^2\sum_{i\neq j} k_i^{\{\nu} k_j^{\mu\}} 
\sum_{l\neq i,j} \frac{(q\theta_j\epsilon_j)(q\theta_l\epsilon_l){I^{(-1)}}^j_j}{(z_i-z_j)(z_j-z_l)}
-\frac{\alpha'}{2}\sum_{i\neq j} \theta_i\epsilon_i^{\{\mu}\,k_i^{\nu\}}\, (q\theta_j\epsilon_j){I^{(0)}}^i_{iij}
\nonumber\\
&&
-\frac{\alpha'}{2}\sum_{i\neq j} \theta_i\epsilon_i^{\{\mu}\,k_j^{\nu\}}\,(q\theta_j\epsilon_j)  {I^{(0)}}^j_{iij}- \frac{\alpha'}{2}\sum_{i\neq j\neq l} \theta_i\epsilon_i^{\{\mu}\,k_j^{\nu\}}\,(q\theta_l\epsilon_l) {I^{(0)}}^j_{iil}\nonumber\\
&&
+\left( \frac{\alpha'}{2}\right)\sum_{i\neq j} \frac{\theta_i\epsilon_i^{\{\mu} (k_i^{\nu\}}+ k_j^{\nu\}})(q\theta_j\epsilon_j)\,}{(z_i-z_j)}{I^{(0)}}^j_{ii}+\left( \frac{\alpha'}{2}\right)\sum_{i\neq j}\frac{(qk_j)\theta_j\epsilon_j^{\{\nu}\,\theta_i\epsilon_i^{\mu\}}}{(z_i-z_j)^2}
\label{B95}
\end{eqnarray}
For the terms involving $I_{i}^j$ we can show that they are matched by part of the first term in $S_1^{(2)}$, by using the following identities:
\begin{eqnarray}
&&I^j_{iij}=I^j_{iji}=\frac{1}{(z_i-z_j)}\left[ I^j_{ii}-\frac{I^j_i-I^j_j}{z_i-z_j}\right]= \frac{{I^{(-1)}}^j_j}{(z_i-z_j)^2}+O(q^0)\nonumber\\
&&I^i_{ilj} = \frac{1}{z_l-z_j}\left[ \frac{I^i_i-I^i_l}{z_i-z_l}-\frac{I^i_i-I^i_j}{z_i-z_j}\right]= \frac{{I^{(-1)}}^i_i}{(z_i-z_j)(z_i-z_l)}+O(q^0)\nonumber\\
&&I^j_{ijl}=I^j_{ilj}= \frac{1}{z_j-z_l}\left[\frac{I^j_i-I^j_j}{z_i-z_j} -\frac{I^j_i-I^j_l}{z_i-z_l}\right]=-\frac{{I^{(-1)}}^j_j}{(z_j-z_l)(z_i-z_j)}+O(q^0)\ .
\end{eqnarray} 
The first term of $S_1^{(2)}$ furthermore contains the following term, unmatched by Eq.~(\ref{B95}):
\begin{eqnarray}
&&\left(\frac{\alpha'}{2}\right)^2 \sum_{i\neq j}  (\epsilon_qk_i)(\bar{\epsilon}_qk_i)(\theta_i\epsilon_iq)(\theta_j\epsilon_jq){I^{(-1)}}^i_{iij}=-\left(\frac{\alpha'}{2}\right)^2 \sum_{i\neq j} \frac{ (\epsilon_qk_i)(\bar{\epsilon}_qk_i)(\theta_i\epsilon_iq)(\theta_j\epsilon_jq)}{(z_i-z_j)^2}{I^{(-1)}}^i_i\nonumber\\
&&\label{rm2}
\end{eqnarray}
The terms involving $I_{iil}^j$ in Eq.~(\ref{B95}) trivially reproduce the first part of $S_3^{(1)}$. Finally, the first and the last line in Eq.~(\ref{B95}) are not present in $S_1$, $S_2$ and $S_3$. 
Again, we discuss on the missing pieces in the end.

Next, we consider
\begin{eqnarray}
&&A^{\mu\nu}_{\theta\bar{\theta}}=\frac{1}{2}\left(\frac{\alpha'}{2}\right)^2\sum_{i\neq j} \frac{(k_i^{\{\mu} k_i^{\nu\}}(\bar{\theta}_j\bar{\epsilon}_jq)+ k_i^{\{\mu}k_j^{\nu\}}(\bar{\theta}_i\bar{\epsilon}_iq) )(q\theta_j\epsilon_j)}{|z_i-z_j|^2}\left( {I^{(-1)}}^i_i+  {I^{(-1)}}^j_j\right)\nonumber\\
&&+\frac{1}{2}\left(\frac{\alpha'}{2}\right)^2 \sum_{i\neq j\neq l}\frac{k_i^{\{\mu} k_i^{\nu\}} (q\theta_j\epsilon_j)(q\bar{\theta}_l\bar{\epsilon}_l)}{(z_i-z_j)(\bar{z}_i-\bar{z}_l)}{I^{(-1)}}^i_i +\frac{1}{2}\left(\frac{\alpha'}{2}\right)^2\sum_{i\neq j\neq l}\frac{k_i^{\{\mu} k_j^{\nu\}} (q\theta_l\epsilon_l)(q\bar{\theta}_l\bar{\epsilon}_l)}{(z_i-z_l)(\bar{z}_j-\bar{z}_l)}{I^{(-1)}}^l_l\nonumber\\
&&+
\left(\frac{\alpha'}{2}\right)^2\sum_{i\neq j\neq l}\frac{k_i^{\{\mu} k_j^{\nu\}} (q\theta_l\epsilon_l)(q\bar{\theta}_i\bar{\epsilon}_i)}{(z_i-z_l)(\bar{z}_i-\bar{z}_j)}{I^{(-1)}}^i_i-\frac{\alpha'}{2}\sum_{i\neq j}  \left(\theta_i\epsilon_i^{\{\mu} k_j^{\nu\}}(\bar{\theta}_i\bar{\epsilon}_iq) +\theta_i\epsilon^{\{\mu} k_i^{\nu\}} (\bar{\theta}_j\bar{\epsilon}_j q)\right){I^{(0)}}_{ii}^{ij}\nonumber\\
&&-\frac{\alpha'}{2}\sum_{i\neq j\neq l}\theta_i\epsilon_i^{\{\mu}k_j^{\nu\}}~(q \bar{\theta}_l\bar{\epsilon}_l) {I^{(0)}}_{ii}^{jl}
+\sum_{i, j=1}^n \theta_i\epsilon_i^{\{\mu} \bar{\theta}_j\bar{\epsilon}_j^{\nu\}} {I^{(1)}}_{ii}^{jj}\label{B113}
\end{eqnarray} 
The first line, the first term of the second line, and the last term are all real, so their complex conjugate simply brings a factor of two. We may use the following identities to show that the first and second line and the first term in the third line match the second part of $S_1^{(2)}$:
\begin{eqnarray}
&&I^{ij}_{ij}=I^{ji}_{ij}= \frac{{I^{(-1)}}^i_i+{I^{(-1)}}^j_j}{|z_i-z_j|^2}+O(q^0)~~;~~I^{il}_{ij}= \frac{{I^{(-1)}}^i_i}{(z_i-z_j)(\bar{z}_i-\bar{z}_l)}+O(q^0)\nonumber\\
&& I^{jl}_{il}=\frac{{I^{(-1)}}^l_l}{(z_i-z_l)(\bar{z}_j-\bar{z}_l)}+O(q^0)~~;~~I^{ij}_{il}=\frac{{I^{(-1)}}^i_i}{(z_i-z_l)(\bar{z}_i-\bar{z}_j)}+O(q^0). 
\end{eqnarray}
The last term in Eq. (\ref{B113}) is trivially matched by $S_2^{(0)}$ and all the remaining terms are matched by the second part of $S_3^{(1)}$. 

Next, consider
\begin{eqnarray}
A_{\theta^3}^{\mu\nu}=  - \left(\frac{\alpha'}{2}\right)^{\frac{3}{2}}\sum_{i\neq j\neq l} \frac{\theta_i\epsilon_i^{\{\mu} ~(q \theta_l\epsilon_l)~(q \theta_j\epsilon_j)}{(z_i-z_j)^2(z_j-z_l)}\Bigg( k_i^{\nu\}} {I^{(-1)}}^i_i
-k_j^{\nu\}} {I^{(-1)}}^j_j\Bigg)\label{ttt}
\end{eqnarray}
It is easily seen that this expression matches the first part of $S_3^{(2)}$ through order $q$, after using the identities
\begin{eqnarray}
&&I^{j}_{iijl}=I^j_{iilj}= \frac{{I^{(-1)}}^j_j}{(z_j-z_l)(z_i-z_j)^2}+O(q^0)\nonumber\\
&&I^{i}_{iijl}= {I^{(-1)}}^i_i\left[-\frac{1}{(z_i-z_j)^2(z_j-z_l)}+\frac{1}{(z_i-z_l)^2(z_j-z_l)}\right]+O(q^0)
\end{eqnarray}
and the property that $I_{iilm}^j$ only contributes at order $q^{-1}$ if either of the indices $i,l$ or $m$ equals $j$. 

Next, we consider
\begin{eqnarray}
&&A^{\mu\nu}_{\theta^2\bar{\theta}}=\sqrt{\frac{\alpha'}{2}} \sum_{i\neq j}
\frac{\bar{\theta}_i\bar{\epsilon}_i^{\{\nu}\big(-
\theta_j\epsilon_j^{\mu\}}~(q \theta_i\epsilon_i)+\theta_i\epsilon_i^{\mu\}}~(q \theta_j\epsilon_j)\big)}{(z_i-z_j)}{I^{(0)}}^{ii}_{ji}
+\sqrt{\frac{\alpha'}{2}}\sum_{i\neq j\neq l} \frac{\bar{\theta}_j\bar{\epsilon}_j^{\{\nu}~\theta_i\epsilon_i^{\mu\}}~ (q \theta_l\epsilon_l)}{(z_i-z_l)} {I^{(0)}}^{jj}_{il}\nonumber\\
&&-\left(\frac{\alpha'}{2}\right)^{\frac{3}{2}}\sum_{i\neq j} \theta_i\epsilon_i^{\{\mu} \left[\frac{k_i^{\nu\}} (q\bar{\theta}_j\epsilon_j) (q\theta_j\epsilon_j)+k_j^{\nu\}} (q\bar{\theta}_i\epsilon_i) (q\theta_j\epsilon_j)}
{|z_i-z_j|^2(z_i-z_j)}\right]\left[{I^{(-1)}}^i_i+{I^{(-1)}}^j_j\right]\nonumber\\
&&-\left(\frac{\alpha'}{2}\right)^{\frac{3}{2}}\sum_{i\neq j \neq l}\theta_i
\epsilon^{\{\mu}\left[\frac{k_i^{\nu\}} (q\bar{\theta}_j\epsilon_j)(q\theta_l\epsilon_l)+k_j^{\nu\}}(q\bar{\theta}_i\bar{\epsilon}_i)(q\theta_l\epsilon_l)}{(z_i-z_l)^2(\bar{z}_i-\bar{z}_j)}\right] {I^{(-1)}}^i_i\nonumber\\
&&+\left(\frac{\alpha'}{2}\right)^{\frac{3}{2}}\sum_{i\neq j\neq l} \frac{\theta_i\epsilon_i^{\{\mu}\,k_j^{\nu\}} (q\bar{\theta}_l\bar{\epsilon}_l)(q\theta_j\epsilon_j)}{(\bar{z}_j-\bar{z}_l)(z_i-z_j)^2} {I^{(-1)}}^j_j-\left(\frac{\alpha'}{2}\right)^{\frac{3}{2}} \sum_{i\neq j\neq l} \frac{\theta_i\epsilon_i^{\{\mu}\,k_j^{\nu\}} (q\bar{\theta}_l\bar{\epsilon}_l)(q\theta_l\epsilon_l)}{(\bar{z}_j-\bar{z}_l)(z_i-z_l)^2}{I^{(-1)}}^l_l\label{ttbt}
\end{eqnarray}
The terms in the first line match $S_2^{(1)}$, since $ {I^{(0)}}^{ii}_{ji}\,(z_i-z_j)^{-1}=-{I^{(0)}}^{ii}_{iij}+ \Ord(q)$ and ${I^{(0)}}^{ii}_{ji}\,(z_i-z_l)^{-1}= -{I^{(0)}}^{jj}_{iil} + \Ord(q)$.
For the remaining terms, all the dependence on the $z_i$ variables in the second line is matched by the integrals ${I^{(-1)}}^{ij}_{iij}={I^{(-1)}}^{ji}_{iij}$, while dependence on the $z_i$ variables in the third line, instead, matches the integrals $ {I^{(-1)}}^{ji}_{iil}={I^{(-1)}}^{ij}_{iil}$. Finally the last two terms follow respectively from the integrals  ${I^{(-1)}}^{jl}_{iij}$      and    ${I^{(-1)}}^{lj}_{iil}$. Altogether these remaining terms thus match the second part of $S_3^{(2)}$. 

The last expression to be considered is
\begin{eqnarray}
&&A^{\mu\nu}_{\theta^2\bar{\theta}^2} =\left(\frac{\alpha'}{2}\right)\sum_{i\neq j}\frac{\theta_i\epsilon_i^{\{\nu}~(q \theta_j\epsilon_j)}{|z_i-z_j|^4}\big(  \bar{\theta}_i\epsilon_i^{\mu\}}~(q \bar{\theta}_j\epsilon_j)-\bar{\theta}_j\epsilon_j^{\mu\}}~(q \bar{\theta}_i\epsilon_i)\big)\Big[{I^{(-1)}}^i_i+{I^{(-1)}}^j_j\Big]\nonumber\\
&&+\left(\frac{\alpha'}{2}\right)\sum_{i\neq j\neq l}\theta_i\epsilon_i^{\{\nu}\Bigg[\frac{(q \theta_j\epsilon_j)~\bar{\theta}_i\bar{\epsilon}_i^{\mu\}}~(q \bar{\theta}_l\epsilon_l)}{(z_i-z_j)^2(\bar{z}_i-\bar{z}_l)^2}{I^{(-1)}}^i_i
-\frac{(q \theta_l\epsilon_l)~\bar{\theta}_j\bar{\epsilon}_j^{\mu\}}~(q
 \bar{\theta}_i\bar{\epsilon}_i)}{(z_i-z_l)^2(\bar{z}_i-\bar{z}_j)^2}{I^{(-1)}}^i_i 
  \nonumber\\
&&
 -\frac{(q \theta_j\epsilon_j)~\bar{\theta}_j\bar{\epsilon}_j^{\mu\}}~(q \bar{\theta}_l\bar{\epsilon}_l)}{(z_i-z_j)^2(\bar{z}_l-\bar{z}_j)^2}{I^{(-1)}}^j_j
+\frac{(q \theta_l\epsilon_l)~\bar{\theta}_j\bar{\epsilon}_j^{\mu\}}~(q \bar{\theta}_l\epsilon_l)}{(z_i-z_l)^2(\bar{z}_l-\bar{z}_j)^2}{I^{(-1)}}^l_l\Bigg]\label{ttbtbt}
\end{eqnarray}
The first line can be expressed in terms of the integrals ${I^{(-1)}}^{iij}_{iij}=-{I^{(-1)}}^{jji}_{iij}$, matching the similar type of terms in $S_2^{(2)}$. Similarly the other terms can be expressed in terms of the integrals ${I^{(-1)}}^{iil}_{iij}$, ${I^{(-1)}}^{jji}_{iil}$, ${I^{(-1)}}^{jjl}_{iij}$ and ${I^{(-1)}}^{jjl}_{iil}$, matching the remaining types of non-vanishing  terms in $S_2^{(2)}$.

Finally, by the same procedures as above, it is easy to check that the remaining terms to be matched in  Eqs.~(\ref{rm1}) and (\ref{rm2}) are obtained by acting on the $n$-point amplitude with the operators defined in the second and third line of Eq.~(\ref{generalsubsub}). These operator simultaneously produce terms, which exactly cancel the unmatched terms in Eqs.~(\ref{theta1}) and (\ref{B95}). This proves the uniqueness of the subsubleading soft operator in Eq.~\eqref{generalsubsub} applicable to the graviton and the dilaton.


\begin{thebibliography}{99}
\fontsize{10pt}{12pt}\selectfont

\bibitem{asymp}
A.~Strominger, 
  \href{http://dx.doi.org/10.1007/JHEP07(2014)151}{{\em JHEP} {\bfseries 07}
  (2014) 151},
\href{http://arxiv.org/abs/1308.0589}{{\ttfamily  arXiv:1308.0589}};\\
%
A.~Strominger, 
  \href{http://dx.doi.org/10.1007/JHEP07(2014)152}{{\em JHEP} {\bfseries 07}
  (2014) 152},
\href{http://arxiv.org/abs/1312.2229}{{\ttfamily  arXiv:1312.2229}};\\
%
T.~He, V.~Lysov, P.~Mitra, and A.~Strominger, 
  \href{http://dx.doi.org/10.1007/JHEP05(2015)151}{{\em JHEP} {\bfseries 05}
  (2015) 151},
\href{http://arxiv.org/abs/1401.7026}{{\ttfamily  arXiv:1401.7026}};\\
%
T.~Adamo, E.~Casali, D.~Skinner, 
\href{http://dx.doi.org/10.1088/0264-9381/31/22/225008}{{\em
  Class.Quant.Grav.} {\bfseries 31} no.~22, (2014) 225008},
\href{http://arxiv.org/abs/1405.5122}{{\ttfamily  arXiv:1405.5122}};\\
%
Y.~Geyer, A.~E. Lipstein, L.~Mason, 
  \href{http://dx.doi.org/10.1088/0264-9381/32/5/055003}{{\em Class.Quant.Grav.} {\bfseries 32} no.~5, (2015) 055003},
\href{http://arxiv.org/abs/1406.1462}{{\ttfamily  arXiv:1406.1462}};\\
%
%
T.~Adamo, E.~Casali, 
  \href{http://dx.doi.org/10.1103/PhysRevD.91.125022}{{\em Phys. Rev.}
  {\bfseries D91} no.~12, (2015) 125022},
\href{http://arxiv.org/abs/1504.02304}{{\ttfamily  arXiv:1504.02304}};\\
%
D.~Kapec, V.~Lysov, S.~Pasterski, and A.~Strominger, 
  \href{http://dx.doi.org/10.1007/JHEP08(2014)058}{{\em JHEP} {\bfseries 08}
  (2014) 058},
\href{http://arxiv.org/abs/1406.3312}{{\ttfamily  arXiv:1406.3312}};\\
%
T.~He, P.~Mitra, A.~P. Porfyriadis, and A.~Strominger, 
\href{http://dx.doi.org/10.1007/JHEP10(2014)112}{{\em JHEP}
  {\bfseries 10} (2014) 112},
\href{http://arxiv.org/abs/1407.3789}{{\ttfamily  arXiv:1407.3789}};\\
%
V.~Lysov, S.~Pasterski, A.~Strominger, 
  \href{http://dx.doi.org/10.1103/PhysRevLett.113.111601}{{\em Phys.Rev.Lett.} {\bfseries 113} no.~11, (2014) 111601},
\href{http://arxiv.org/abs/1407.3814}{{\ttfamily  arXiv:1407.3814}};\\
%
M.~Campiglia and A.~Laddha, 
\href{http://dx.doi.org/10.1103/PhysRevD.90.124028}{{\em
  Phys. Rev.} {\bfseries D90} no.~12, (2014) 124028},
\href{http://arxiv.org/abs/1408.2228}{{\ttfamily  arXiv:1408.2228}};\\
%
D.~Kapec, V.~Lysov, and A.~Strominger, 
\href{http://arxiv.org/abs/1412.2763}{{\ttfamily  arXiv:1412.2763}};\\
%
A.~Mohd, 
  \href{http://dx.doi.org/10.1007/JHEP02(2015)060}{{\em JHEP} {\bfseries 02}
  (2015) 060},
\href{http://arxiv.org/abs/1412.5365}{{\ttfamily  arXiv:1412.5365}};\\
%
M.~Campiglia and A.~Laddha, 
\href{http://dx.doi.org/10.1007/JHEP04(2015)076}{{\em JHEP}
  {\bfseries 04} (2015) 076},
\href{http://arxiv.org/abs/1502.02318}{{\ttfamily  arXiv:1502.02318}};\\
%
D.~Kapec, V.~Lysov, S.~Pasterski, and A.~Strominger, 
\href{http://arxiv.org/abs/1502.07644}{{\ttfamily  arXiv:1502.07644}};\\
%
T.~He, P.~Mitra, and A.~Strominger, 
\href{http://arxiv.org/abs/1503.02663}{{\ttfamily  arXiv:1503.02663}};\\
%
M.~Campiglia and A.~Laddha, 
\href{http://dx.doi.org/10.1007/JHEP07(2015)115}{{\em
  JHEP} {\bfseries 07} (2015) 115},
\href{http://arxiv.org/abs/1505.05346}{{\ttfamily  arXiv:1505.05346}};\\
%
D.~Kapec, M.~Pate, and A.~Strominger, 
\href{http://arxiv.org/abs/1506.02906}{{\ttfamily  arXiv:1506.02906}};\\
%
S.~G. Avery and B.~U.~W. Schwab, 
\href{http://dx.doi.org/10.1103/PhysRevD.93.026003
}{{\em Phys. Rev.} {\bfseries D93}
  (2016) 026003},
\href{http://arxiv.org/abs/1506.05789}{{\ttfamily  arXiv:1506.05789}};\\
%
M.~Campiglia and A.~Laddha, 
\href{http://dx.doi.org/10.1007/JHEP12(2015)094}{{\em JHEP} {\bfseries 12}
  (2015) 094},
\href{http://arxiv.org/abs/1509.01406}{{\ttfamily  arXiv:1509.01406}};\\
%
S.~G. Avery and B.~U.~W. Schwab, 
\href{http://dx.doi.org/10.1007/JHEP02(2016)031}{{\em JHEP} {\bfseries 02}
  (2016) 031},
\href{http://arxiv.org/abs/1510.07038}{{\ttfamily  arXiv:1510.07038}};\\
%
%
  T.~T.~Dumitrescu, T.~He, P.~Mitra and A.~Strominger,
\href{http://arxiv.org/abs/1511.07429}{{\ttfamily   arXiv:1511.07429}};\\
%
%
  M.~Mirbabayi and M.~Simonovic,
  \href{http://arxiv.org/abs/1602.05196}{{\ttfamily   arXiv:1602.05196}}.


\bibitem{Boels:2015pta}
R. Boels and W. Wormsbecher,
\href{http://arxiv.org/abs/arXiv:1507.08162}{{\ttfamily arXiv:1507.08162}.}

\bibitem{DiVecchia:2015jaq} 
  P.~Di Vecchia, R.~Marotta, M.~Mojaza and J.~Nohle,
\href{http://arxiv.org/abs/ arXiv:1512.03316}{{\ttfamily arXiv:1512.03316}.} 
  
\bibitem{Cheung:2014dqa} 
  C.~Cheung, K.~Kampf, J.~Novotny and J.~Trnka,
\href{http://dx.doi.org/10.1103/PhysRevLett.114.221602}{{\em  Phys.\ Rev.\ Lett.}  {\bf 114}, no. 22, 221602 (2015)}
\href{http://arxiv.org/abs/ arXiv:1412.4095}{{\ttfamily arXiv:1412.4095}.}

\bibitem{Luo:2015tat} 
  H.~Luo and C.~Wen,
  \href{http://dx.doi.org/10.1007/JHEP03(2016)088}{{\em JHEP} {\bf 1603}, 088 (2016)}
\href{http://arxiv.org/abs/ arXiv:1512.06801}{{\ttfamily arXiv:1512.06801}.}


\bibitem{Low} 
F.~E.~Low,
\href{http://dx.doi.org/10.1103/PhysRev.96.1428}{{\em
Phys.\ Rev.\ }  {\bf 96}, 1428 (1954)};\\
%
M.~Gell-Mann and M.~L.~Goldberger,
\href{http://dx.doi.org/ 10.1103/PhysRev.96.1433}{{\em 
Phys.\ Rev.}\  {\bf 96}, 1433 (1954)};\\
%
F.~E.~Low,
 \href{http://dx.doi.org/10.1103/PhysRev.110.974}{{\em 
Phys.\ Rev.}\  {\bf 110}, 974 (1958)};\\
See also:\\
S.~Saito,
 \href{http://dx.doi.org/10.1103/PhysRev.184.1894}{{\em 
Phys.\ Rev.}\  {\bf 184}, 1894 (1969)};\\
T.~H.~Burnett and N.~M.~Kroll,
\href{http://dx.doi.org/10.1103/PhysRevLett.20.86}{{\em 
  Phys.\ Rev.\ Lett.}\  {\bf 20}, 86 (1968)};\\
%
J.~S.~Bell and R.~Van Royen,
\href{http://dx.doi.org/10.1007/BF02823297}{{\em 
  Nuovo Cim.}\  {\bf A60}, 62 (1969)};\\
%
 V.~Del Duca,
\href{http://dx.doi.org/ 10.1016/0550-3213(90)90392-Q}{{\em 
  Nucl.\ Phys.}\  {\bf B345}, 369 (1990)}.


\bibitem{Weinberg}
S.~Weinberg,
\href{http://dx.doi.org/10.1103/PhysRev.135.B1049}{{\em 
Phys.\ Rev.}\  {\bf 135}, B1049 (1964)};\\
%
S.~Weinberg,
\href{http://dx.doi.org/10.1103/PhysRev.140.B516}{{\em 
Phys.\ Rev.}\  {\bf 140}, B516 (1965)}; \\
D.~J.~Gross and R.~Jackiw,
\href{http://dx.doi.org/10.1103/PhysRev.166.1287}{{\em 
Phys.\ Rev.}\  {\bf 166}, 1287 (1968)};\\
R.~Jackiw,
\href{http://dx.doi.org/10.1103/PhysRev.168.1623}{{\em 
 Phys.\ Rev.}\  {\bf 168}, 1623 (1968)}.

\bibitem{GenericSubStart}
E.~Laenen, G.~Stavenga and C.~D.~White,
\href{http://dx.doi.org/ 10.1088/1126-6708/2009/03/054}{{\em 
JHEP} {\bf 0903},  (2009) 054},
\href{http://arxiv.org/abs/arXiv:0811.2067}{{\ttfamily arXiv:0811.2067};} \\
%
E.~Laenen, L.~Magnea, G.~Stavenga and C.~D.~White,
\href{http://dx.doi.org/ 10.1007/JHEP01(2011)141}{{\em 
JHEP} {\bf 1101}, 141 (2011)}, 
\href{http://arxiv.org/abs/arXiv:1010.1860}{{\ttfamily arXiv:1010.1860} ;}\\
C.~D.~White,
\href{http://dx.doi.org/10.1007/JHEP05(2011)060}{{\em 
JHEP} {\bf 1105},  (2011) 060}, 
\href{http://arxiv.org/abs/arXiv:1103.2981}{{\ttfamily arXiv:1103.2981}.}





\bibitem{SoftGravityYangMills}
F.~Cachazo and A.~Strominger, 
\href{http://arxiv.org/abs/1404.4091}{{\ttfamily  arXiv:1404.4091}};\\
%
E.~Casali, 
  \href{http://dx.doi.org/10.1007/JHEP08(2014)077}{{\em JHEP} {\bfseries 08}
  (2014) 077},
\href{http://arxiv.org/abs/1404.5551}{{\ttfamily  arXiv:1404.5551}};\\
%
B.~U.~W. Schwab and A.~Volovich, 
  \href{http://dx.doi.org/10.1103/PhysRevLett.113.101601}{{\em Phys. Rev.
  Lett.} {\bfseries 113} no.~10, (2014) 101601},
\href{http://arxiv.org/abs/1404.7749}{{\ttfamily  arXiv:1404.7749}};\\
%
Z.~Bern, S.~Davies, and J.~Nohle, 
  \href{http://dx.doi.org/10.1103/PhysRevD.90.085015}{{\em Phys. Rev.}
  {\bfseries D90} no.~8, (2014) 085015},
\href{http://arxiv.org/abs/1405.1015}{{\ttfamily  arXiv:1405.1015}};\\
%
S.~He, Y.-t. Huang, and C.~Wen, 
  \href{http://dx.doi.org/10.1007/JHEP12(2014)115}{{\em JHEP} {\bfseries 12}
  (2014) 115},
\href{http://arxiv.org/abs/1405.1410}{{\ttfamily  arXiv:1405.1410}};\\
%
A.~J. Larkoski, 
\href{http://dx.doi.org/10.1103/PhysRevD.90.087701}{{\em Phys.
  Rev.} {\bfseries D90} no.~8, (2014) 087701},
\href{http://arxiv.org/abs/1405.2346}{{\ttfamily  arXiv:1405.2346}};\\
%
F.~Cachazo and E.~Y. Yuan, 
\href{http://arxiv.org/abs/1405.3413}{{\ttfamily  arXiv:1405.3413}};\\
%
N.~Afkhami-Jeddi, 
\href{http://arxiv.org/abs/1405.3533}{{\ttfamily  arXiv:1405.3533}};\\
%
J.~Broedel, M.~de~Leeuw, J.~Plefka, M.~Rosso, 
  \href{http://dx.doi.org/10.1103/PhysRevD.90.065024}{{\em Phys.Rev.}
  {\bfseries D90} no.~6, (2014) 065024},
\href{http://arxiv.org/abs/1406.6574}{{\ttfamily  arXiv:1406.6574}};\\
%
C.~D. White, 
  \href{http://dx.doi.org/10.1016/j.physletb.2014.08.041}{{\em Phys. Lett.}
  {\bfseries B737} (2014) 216--222},
\href{http://arxiv.org/abs/1406.7184}{{\ttfamily  arXiv:1406.7184}};\\
M.~Zlotnikov, 
  \href{http://dx.doi.org/10.1007/JHEP10(2014)148}{{\em JHEP}
  {\bfseries 10} (2014) 148},
\href{http://arxiv.org/abs/1407.5936}{{\ttfamily  arXiv:1407.5936}};\\
%
C.~Kalousios and F.~Rojas, 
  \href{http://dx.doi.org/10.1007/JHEP01(2015)107}{{\em JHEP} {\bfseries 01}
  (2015) 107},
\href{http://arxiv.org/abs/1407.5982}{{\ttfamily  arXiv:1407.5982}};\\
%
Y.-J. Du, B.~Feng, C.-H. Fu, and Y.~Wang, 
\href{http://dx.doi.org/10.1007/JHEP11(2014)090}{{\em JHEP}
  {\bfseries 11} (2014) 090},
\href{http://arxiv.org/abs/1408.4179}{{\ttfamily  arXiv:1408.4179}};\\
%
D.~Bonocore, E.~Laenen, L.~Magnea, L.~Vernazza, and C.~D. White, 
\mbox{\href{http://dx.doi.org/10.1016/j.physletb.2015.02.008}{{\em Phys. Lett.}
  {\bfseries B742} (2015) 375--382},}
\href{http://arxiv.org/abs/1410.6406}{{\ttfamily  arXiv:1410.6406}};\\
%
H.~Luo, P.~Mastrolia, W.~J. Torres~Bobadilla, 
  \href{http://dx.doi.org/10.1103/PhysRevD.91.065018}{{\em Phys.Rev.}
  {\bfseries D91} no.~6, (2015) 065018},
\href{http://arxiv.org/abs/1411.1669}{{\ttfamily  arXiv:1411.1669}};\\
%
J.~Broedel, M.~de~Leeuw, J.~Plefka, M.~Rosso, 
  \href{http://dx.doi.org/10.1016/j.physletb.2015.05.018}{{\em Phys.Lett.}
  {\bfseries B746} (2015) 293--299},
\href{http://arxiv.org/abs/1411.2230}{{\ttfamily  arXiv:1411.2230}};\\
%
A.~J. Larkoski, D.~Neill, and I.~W. Stewart, 
\href{http://dx.doi.org/10.1007/JHEP06(2015)077}{{\em JHEP}
  {\bfseries 06} (2015) 077},
\href{http://arxiv.org/abs/1412.3108}{{\ttfamily  arXiv:1412.3108}};\\
%
A.~Sabio~Vera and M.~A. Vazquez-Mozo, 
\href{http://dx.doi.org/10.1007/JHEP03(2015)070}{{\em JHEP}
  {\bfseries 03} (2015) 070},
\href{http://arxiv.org/abs/1412.3699}{{\ttfamily  arXiv:1412.3699}};\\
%
A.~E. Lipstein, 
  \href{http://dx.doi.org/10.1007/JHEP06(2015)166}{{\em JHEP} {\bfseries 06}
  (2015) 166},
\href{http://arxiv.org/abs/1504.01364}{{\ttfamily  arXiv:1504.01364}};\\
%
S.~D. Alston, D.~C. Dunbar, and W.~B. Perkins, 
  \href{http://dx.doi.org/10.1103/PhysRevD.92.065024}{{\em Phys. Rev.}
  {\bfseries D92} no.~6, (2015) 065024},
\href{http://arxiv.org/abs/1507.08882}{{\ttfamily  arXiv:1507.08882}};\\
%
%
  J.~Rao and B.~Feng,
  \href{http://arxiv.org/abs/1604.00650}{{\ttfamily  arXiv:1604.00650}}.


\bibitem{BDDN}
Z. Bern, S. Davies, P. Di Vecchia and J. Nohle, 
\href{http://dx.doi.org/10.1103/PhysRevD.90.084035}{{\em Phys. Rev. }{\bf D90} (2014) 8, 084035}, 
\href{http://arxiv.org/abs/arXiv:1406.6987}{{\ttfamily arXiv:1406.6987}.} 

  \bibitem{BianchiR2phi}
M.~Bianchi, S.~He, Y.-t. Huang, and C.~Wen, 
  \href{http://dx.doi.org/10.1103/PhysRevD.92.065022}{{\em Phys. Rev.}
  {\bfseries D92} no.~6, (2015) 065022},
\href{http://arxiv.org/abs/1406.5155}{{\ttfamily  arXiv:1406.5155}};

\bibitem{softsusy}
Z.-W. Liu, 
  \href{http://dx.doi.org/10.1140/epjc/s10052-015-3304-1}{{\em Eur. Phys. J.}
  {\bfseries C75} no.~3, (2015) 105},
\href{http://arxiv.org/abs/1410.1616}{{\ttfamily  arXiv:1410.1616}};\\
%
J.~Rao, 
\href{http://dx.doi.org/10.1007/JHEP02(2015)087}{{\em JHEP}
  {\bfseries 02} (2015) 087},
\href{http://arxiv.org/abs/1410.5047}{{\ttfamily  arXiv:1410.5047}};\\
%
W.-M. Chen, Y.-t. Huang, and C.~Wen, 
  \href{http://dx.doi.org/10.1103/PhysRevLett.115.021603}{{\em Phys.Rev.Lett.} {\bfseries 115} no.~2, (2015) 021603},
\href{http://arxiv.org/abs/1412.1809}{{\ttfamily  arXiv:1412.1809}};\\
%
W.-M. Chen, Y.-t. Huang, and C.~Wen, 
  \href{http://dx.doi.org/10.1007/JHEP03(2015)150}{{\em JHEP} {\bfseries 03}
  (2015) 150},
\href{http://arxiv.org/abs/1412.1811}{{\ttfamily  arXiv:1412.1811}};\\
%
L.~V. Bork and A.~I. Onishchenko, 
\href{http://dx.doi.org/10.1007/JHEP12(2015)030}{{\em JHEP} {\bfseries 12}
  (2015) 030},
\href{http://arxiv.org/abs/1506.07551}{{\ttfamily  arXiv:1506.07551}};\\
%
S.~Chin, S.~Lee, and Y.~Yun, 
\href{http://dx.doi.org/10.1007/ 10.1007/JHEP11(2015)088}{{\em JHEP} {\bfseries 11}
  (2015) 088},
\href{http://arxiv.org/abs/1508.07975}{{\ttfamily  arXiv:1508.07975}};\\
%
A.~Brandhuber, E.~Hughes, B.~Spence, and G.~Travaglini, 
\href{http://dx.doi.org/10.1007/JHEP03(2016)084}{{\em JHEP} {\bfseries 03}
  (2016) 084},
\href{http://arxiv.org/abs/1511.06716}{{\ttfamily  arXiv:1511.06716}};\\
%
  S.~G.~Avery and B.~U.~W.~Schwab,
  \href{http://arxiv.org/abs/1512.02657}{ {\ttfamily arXiv:1512.02657}};\\ 
  %
  %
  V.~Lysov,
  \href{http://arxiv.org/abs/1512.03015}{ {\ttfamily arXiv:1512.03015}}.






\bibitem{softstring}
B.~U.~W. Schwab, 
  \href{http://dx.doi.org/10.1007/JHEP08(2014)062}{{\em JHEP} {\bfseries 08}
  (2014) 062},
\href{http://arxiv.org/abs/1406.4172}{{\ttfamily  arXiv:1406.4172}};\\
%
B.~U.~W. Schwab, 
  \href{http://dx.doi.org/10.1007/JHEP03(2015)140}{{\em JHEP} {\bfseries 03}
  (2015) 140},
\href{http://arxiv.org/abs/1411.6661}{{\ttfamily  arXiv:1411.6661}};\\
%
M.~Bianchi and A.~L. Guerrieri, 
  \href{http://dx.doi.org/10.1007/JHEP09(2015)164}{{\em JHEP} {\bfseries 09}
  (2015) 164},
\href{http://arxiv.org/abs/1505.05854}{{\ttfamily  arXiv:1505.05854}};\\
A.~L. Guerrieri,
IFAE 2015, Rome, Italy, April 8-10, 2015,
\href{http://arxiv.org/abs/1507.08829}{{\ttfamily  arXiv:1507.08829}};\\
%
P.~Di~Vecchia, R.~Marotta, and M.~Mojaza, 
\href{http://dx.doi.org/10.1002/prop201500068}{{\em Fortschr. Phys.} {\bfseries 64}
  (2016) 389},
\href{http://arxiv.org/abs/1511.04921}{{\ttfamily  arXiv:1511.04921}}.




\bibitem{DiVecchia:2015oba} 
  P.~Di Vecchia, R.~Marotta and M.~Mojaza,
 \href{http://dx.doi.org/ 10.1007/JHEP05(2015)137}{{\em 
  JHEP} {\bf 1505}, 137 (2015)},
\href{http://arxiv.org/abs/arXiv:1502.05258}{{\ttfamily arXiv:1502.05258}.} 
  


\bibitem{DoubleSoft}
F.~Cachazo, S.~He, and E.~Y. Yuan, 
  \href{http://dx.doi.org/10.1103/PhysRevD.92.065030}{{\em Phys. Rev.}
  {\bfseries D92} no.~6, (2015) 065030},
\href{http://arxiv.org/abs/1503.04816}{{\ttfamily  arXiv:1503.04816}};\\
%
T.~Klose, T.~McLoughlin, D.~Nandan, J.~Plefka, G.~Travaglini,
  \\
  \href{http://dx.doi.org/10.1007/JHEP07(2015)135}{{\em JHEP} {\bfseries 07}
  (2015) 135},
\href{http://arxiv.org/abs/1504.05558}{{\ttfamily  arXiv:1504.05558}};\\
%
A.~Volovich, C.~Wen, and M.~Zlotnikov, 
\href{http://dx.doi.org/10.1007/JHEP07(2015)095}{{\em
  JHEP} {\bfseries 07} (2015) 095},
\href{http://arxiv.org/abs/1504.05559}{{\ttfamily  arXiv:1504.05559}};\\
%
P.~Di~Vecchia, R.~Marotta, and M.~Mojaza, 
\href{http://dx.doi.org/10.1007/JHEP12(2015)150}{{\em JHEP} {\bfseries 12} (2015) 150},
\href{http://arxiv.org/abs/1507.00938}{{\ttfamily  arXiv:1507.00938}};\\
%
  I.~Low,
  \href{http://dx.doi.org/10.1103/PhysRevD.93.045032}{{\em  Phys.Rev.}  {\bfseries D93}  no.4 (2016) 045032 },
\href{http://arxiv.org/abs/1512.01232}{{\ttfamily   arXiv:1512.01232}}.
  
  
  
  \bibitem{ArkaniHamed:2008gz}
N. Arkani-Hamed, F. Cachazo and J. Kaplan,
\href{http://dx.doi.org/10.1007/JHEP09(2010)016}{{\em 
JHEP} {\bf 1009} (2010) 016},
\href{http://arxiv.org/abs/0808.1446}{{\ttfamily arXiv:0808.1446.}} 





\bibitem{Ademollo:1975pf}
M. Ademollo, A. D'Adda, R. D'Auria, F. Gliozzi, E. Napolitano,  S. Sciuto and 
P. Di Vecchia,
\mbox{\href{http://dx.doi.org/10.1016/0550-3213(75)90491-5}{{\em 
Nucl.\ Phys.\ } {\bf B94}, (1975) 221}.}

\bibitem{Shapiro:1975cz}
J. Shapiro,  
\href{http://dx.doi.org/ 10.1103/PhysRevD.11.2937}{{\em Phys.
Rev.} {\bf D11} (1975) 2937}.

\bibitem{YoneyaHata}
T. Yoneya, 
\href{http://dx.doi.org/ 10.1016/0370-2693(87)90345-5}{{\em 
Phys. Lett. }{\bf B 197} (1987) 76};\\
H. Hata,
\href{http://dx.doi.org/10.1143/PTP.88.1197}{{\em 
Progr. Theor. Phys.} {\bf  88} (1992) 1197}.




\bibitem{Metsaev:1987zx} 
  R.~R.~Metsaev and A.~A.~Tseytlin,
  \href{http://dx.doi.org/10.1016/0550-3213(87)90077-0}
 {{\em Nucl.\ Phys.}\  {\bf B293}, 385 (1987)}.
\bibitem{1512.00803}
M. Bianchi, A.  L. Guerrieri,   
\href{http://dx.doi.org/ 10.1016/j.nuclphysb.2016.02.005}{{\em Nucl. Phys.}
 {\bf B905}, 188 (2016)},\href{http://arxiv.org/abs/arXiv:1512.00803}{{\ttfamily arXiv:1512.00803.}} 

\bibitem{Zwiebach:1985uq}
B. Zwiebach 
\href{http://dx.doi.org/10.1016/0370-2693(85)91616-8}
{{\em Phys. Lett.} {\bf B  156} (1985)  315}.







\end{thebibliography}
\end{document}